\documentclass[aps,twocolumn,superscriptaddress]{revtex4-1}
\usepackage[dvipdfmx]{graphicx}

\begin{document}

\title{Spectrum of the Dicke model in a superconducting qubit-oscillator system}

\author{S. Ashhab}
\affiliation{Qatar Environment and Energy Research Institute, Hamad Bin Khalifa University, Qatar Foundation, Doha, Qatar}
\author{Y. Matsuzaki}
\altaffiliation{Present address: Electronics and Photonics Research Institute, National Institute of Advanced Industrial Science and Technology (AIST),Tsukuba, Ibaraki 305-8565, Japan}
\affiliation{NTT Basic Research Laboratories, NTT Corporation, 3-1 Morinosato-Wakamiya, Atsugi, Kanagawa 243-0198, Japan}
\author{K. Kakuyanagi}
\affiliation{NTT Basic Research Laboratories, NTT Corporation, 3-1 Morinosato-Wakamiya, Atsugi, Kanagawa 243-0198, Japan}
\author{S. Saito}
\affiliation{NTT Basic Research Laboratories, NTT Corporation, 3-1 Morinosato-Wakamiya, Atsugi, Kanagawa 243-0198, Japan}
\author{F. Yoshihara}
\affiliation{Advanced ICT Institute, National Institute of Information and Communications Technology, 4-2-1, Nukuikitamachi, Koganei, Tokyo 184-8795, Japan}
\author{T. Fuse}
\affiliation{Advanced ICT Institute, National Institute of Information and Communications Technology, 4-2-1, Nukuikitamachi, Koganei, Tokyo 184-8795, Japan}
\author{K. Semba}
\affiliation{Advanced ICT Institute, National Institute of Information and Communications Technology, 4-2-1, Nukuikitamachi, Koganei, Tokyo 184-8795, Japan}

\begin{abstract}
We calculate the transmission spectrum of a superconducting circuit realization of the Dicke model and identify spectroscopic features that can serve as signatures of the superradiant phase. In particular, we calculate the resonance frequencies of the system as functions of the bias term, which is usually absent in studies on the Dicke model but is commonly present in superconducting qubit circuits. To avoid over-complicating the proposed circuit, we assume a fixed coupling strength. This situation precludes the possibility of observing signatures of the phase transition by varying the coupling strength across the critical point. We show that the spectrum obtained by varying the bias point under fixed coupling strength can contain signatures of the normal and superradiant phases: in the normal phase one expects to observe two spectral lines, while in the superradiant phase four spectral lines are expected to exist close to the qubits' symmetry point. Provided that parameter fluctuations and decoherence rates are sufficiently small, the four spectral lines should be observable and can serve as a signature of the superradiant phase.
\end{abstract}

\maketitle

\section{Introduction}
\label{Sec:Introduction}

The field of cavity quantum electrodynamics (QED), in which one or more atoms interact with the electromagnetic field inside a cavity, has been used as a model for studying light-matter interaction at the fundamental level and has been used to develop a number of technologies over the past several decades \cite{QuantumOpticsBooks}.

The collective interaction of an ensemble of atoms with a cavity field, as described by the Dicke model, gives rise to a number of interesting phenomena. One of these is superradiance, in which the atoms exhibit an accelerated emission of photons into the cavity compared to what one might expect from treating the atoms as independent emitters \cite{Dicke}. Another interesting phenomenon is the occurrence of a phase transition and the emergence of strongly correlated atom-cavity states when the atom-cavity coupling strength exceeds a certain critical value \citep{Hepp,Wang1973,EmaryPRL,EmaryPRE,Bastidas,Dasgupta}. The situation with strongly correlated states is sometimes called the superradiant phase, although it should be emphasized that in this context superradiant states do not exhibit superradiance in the sense of emitting radiation that propagates out of the system. In contrast, when the atom-cavity coupling is weak, the ground state of the atom-cavity system is one in which the individual atoms and the cavity are to a good approximation in their respective ground states with little correlation between them. In this case the system is said to be in the normal phase. Alternatively, one could say that the normal phase is associated with a single dynamically stable state, namely the ground state, while the superradiant phase has multiple macroscopically distinct dynamically stable states.

For several decades, studies on cavity-QED were limited to very small values of the coupling strength. The recent development of circuit-QED using superconducting circuits has led to remarkable advances in the field of cavity-QED. Among these advances is the demonstration of ultrastrong and deep-strong coupling between a single superconducting qubit and a superconducting resonator \cite{Niemczyk,FornDiaz2010,FornDiaz2016,YoshiharaDSC,FornDiazReview,KockumReview}. Superradiance was also observed recently in a circuit-QED system \cite{Mlynek}. There have also been several recent experiments on superconducting quantum metamaterials involving large numbers of qubits or resonators \cite{Shapiro,Anderson,Kakuyanagi,Kollar,Wang2018}.

One of the important tools in studying circuit-QED systems is spectroscopy, in which a probe signal is sent towards the circuit and the reflected and/or transmitted signal gives information about the energy level spacings or frequencies of oscillation modes in the circuit. For example, spectroscopy was used in Refs.~\cite{Wallraff,Niemczyk,FornDiaz2010,FornDiaz2016,YoshiharaDSC,YoshiharaSpectra,YoshiharaLamb} to demonstrate the realization of various strong coupling regimes in circuit-QED systems. It was also used in Ref.~\cite{Kakuyanagi} to quantify the coupling strength between a qubit ensemble and a superconducting resonator. In this work we investigate the spectral features that could be used as signatures of the superradiant phase in a circuit-QED realization of the Dicke model. In particular, we include a finite bias term in the Hamiltonian of the circuit \cite{EmaryPRA}. This term is commonly present in superconducting qubit circuits, especially those involving flux qubits, whereas it is generally not included in conventional cavity-QED studies. This term is usually the easiest parameter to vary, and spectra are often plotted with the bias parameter being one of the variable parameters in the spectra. It is also common in theoretical studies to investigate changes in the system as the coupling strength is varied. Tuning of the coupling strength generally requires rather complicated circuitry that has not been used in ultrastrong or deep-strong coupling circuit-QED experiments to date. We therefore focus on the case where the coupling strength is fixed. We then look for signatures of the superradiant phase in spectra where the only variable parameter is the qubit bias parameter. We find that such signatures do indeed exist, most notably in the appearance of additional spectral lines in the superradiant phase. We assess the feasibility of observing these spectra in superconducting circuits with realistic parameters.

The remainder of this paper is organized as follows: In Sec.~\ref{Sec:Model} we briefly introduce the Dicke model and discuss its stable states, including the ground state and metastable excited state. In Sec.~\ref{Sec:Spectrum} we analyze the spectra that are expected in the normal and superradiant phases of the Dicke model. In Sec.~\ref{Sec:ParameterFluctuations} we discuss the effect of fluctuations in the qubit parameters on the spectra. In Sec.~\ref{Sec:ExperimentalConsiderations} we discuss the possibility of observing the predicted spectra in a typical experimental setup. We finally give some concluding remarks in Sec.~\ref{Sec:Conclusion}.

\section{Ground state of the Dicke model in the absence of fluctuations}
\label{Sec:Model}

We consider a system described by the Dicke model, i.e.~$N$ qubits coupled to a single harmonic oscillator. We first assume that the qubit parameters are identical for all the qubits. The Hamiltonian of this system is given by
\begin{widetext}
\begin{equation}
\hat{H} = \frac{\Delta}{2} \sum_{i=1}^N \hat{\sigma}_z^{(i)} + \frac{\epsilon}{2} \sum_{i=1}^N \hat{\sigma}_x^{(i)} + \hbar\omega \left( \hat{a}^\dagger \hat{a} + \frac{1}{2} \right) + g \sum_{i=1}^N \hat{\sigma}_x^{(i)} (\hat{a}+\hat{a}^\dagger),
\end{equation}
\end{widetext}
where $\Delta$ is the qubit gap, $\epsilon$ is the qubit bias parameter, $\omega$ is the cavity's characteristic frequency, $g$ is the coupling strength between a single qubit and the cavity, the operators $\hat{\sigma}_{\alpha}^{(i)}$ (with $\alpha=x,y,z$) are the Pauli operators of qubit $i$, and $\hat{a}$ and $\hat{a}^{\dagger}$ are, respectively, the annihilation and creation operators of the cavity. The signs in this Hamiltonian are one possible combination out of several other equivalent ones.

In the absence of parameter fluctuations, it is natural to define the collective spin operators
\begin{equation}
\hat{S}_{\alpha}=\sum_{i=1}^N \frac{\hat{\sigma}_{\alpha}^{(i)}}{2},
\end{equation}
which obey the standard spin commutation relations up to the factor $\hbar$, which we have not included in the definition of $\hat{S}_{\alpha}$, i.e.~$\left[\hat{S}_{\alpha},\hat{S}_{\beta}\right]=i\varepsilon_{\alpha\beta\gamma} \hat{S}_{\gamma}$, where $\varepsilon_{\alpha\beta\gamma}$ is the Levi-Civita tensor. If we also define the operators
\begin{equation}
\hat{x} = \frac{ \hat{a} + \hat{a}^\dagger }{2}
\end{equation}
and
\begin{equation}
\hat{p}_x = \frac{ \hat{a} - \hat{a}^\dagger }{2i},
\end{equation}
the Hamiltonian can be expressed as
\begin{eqnarray}
\hat{H} & = & \Delta \hat{S}_z + \epsilon \hat{S}_x + \frac{\hbar\omega}{4} \left( \hat{a} + \hat{a}^\dagger \right)^2 - \frac{\hbar\omega}{4} \left( \hat{a} - \hat{a}^\dagger \right)^2 \nonumber \\ & & + 2 g \hat{S}_x (\hat{a}+\hat{a}^\dagger) \nonumber \\
& = & \Delta \hat{S}_z + \epsilon \hat{S}_x + \hbar\omega \hat{x}^2 + \hbar\omega \hat{p}_x^2 + 4 g \hat{S}_x \hat{x}.
\label{Eq:HamiltonianUsingCollectiveVariables}
\end{eqnarray}
If we now take the classical limit, i.e. treat the spin $S$ as a continuous classical variable (with $\sqrt{S_x^2+S_y^2+S_z^2}=N/2$) and similarly treat $x$ and $p$ as classical position and momentum variables, and we look for the ground state of the system by minimizing the Hamiltonian, we find that this state obeys the relations
\begin{eqnarray}
p_x & = & 0 \nonumber \\
S_z & = & - \frac{N}{2} \cos \theta \nonumber \\
S_x & = & - \frac{N}{2} \sin \theta \nonumber \\
\theta & = & \tan^{-1} \frac{\epsilon+4gx}{\Delta} \nonumber \\
x & = & - \frac{2gS_x}{\hbar\omega}.
\label{Eq:ClassicalGroundState}
\end{eqnarray}
Combining these equations we obtain
\begin{equation}
x = \frac{2g}{\hbar\omega} \times \frac{N}{2} \times \frac{\epsilon+4gx}{\sqrt{\Delta^2+(\epsilon+4gx)^2}}.
\label{Eq:ClassicalGroundStateX}
\end{equation}
For $4g^2N/(\hbar\omega\Delta)<1$, this equation always has a single solution, regardless of the value of $\epsilon$. In the special case $\epsilon=0$, the solution is $x=0$. For finite values of $\epsilon$, the equation becomes less amenable to algebraic manipulation, but a numerical solution can be obtained straightforwardly. When $4g^2N/(\hbar\omega\Delta)>1$ and $\epsilon=0$, Eq.~(\ref{Eq:ClassicalGroundStateX}) has three solutions: $x=0$ (which now is a local maximum of the energy and therefore does not correspond to the ground state) and
\begin{equation}
1 = \frac{4g^2N}{\hbar\omega\Delta} \frac{1}{\sqrt{1+(4gx/\Delta)^2}},
\end{equation}
or in other words
\begin{equation}
x = \pm \frac{\Delta}{4g} \sqrt{\left(\frac{4g^2N}{\hbar\omega\Delta}\right)^2 - 1}.
\label{Eq:CavityFieldInSuperradiantState}
\end{equation}
The plus and minus signs correspond to two equivalent ground states, keeping in mind that we are taking the classical limit and that quantum effects would hybridize the classical solutions and lift the degeneracy. When $4g^2N/(\hbar\omega\Delta)>1$ and $\epsilon$ is finite but small, Eq.~(\ref{Eq:ClassicalGroundStateX}) still has three solutions, but no simple expressions can be derived for them. Two of these solutions, specifically the smallest and the largest values of $x$, correspond to dynamically stable states of the system, while the middle one corresponds to a dynamically unstable state. One of the two stable states is the ground state and the other is a metastable excited state. For sufficiently large values of $|\epsilon|$, two of the solutions disappear and Eq.~(\ref{Eq:ClassicalGroundStateX}) again has a single solution. In this regime there is no metastable excited state any more. Hence for any given value of $4g^2N/(\hbar\omega\Delta)$ above the critical value 1, there is a critical value of $|\epsilon|$ that separates the regions of one and three solutions.

\begin{figure}[h]
\includegraphics[width=6.0cm]{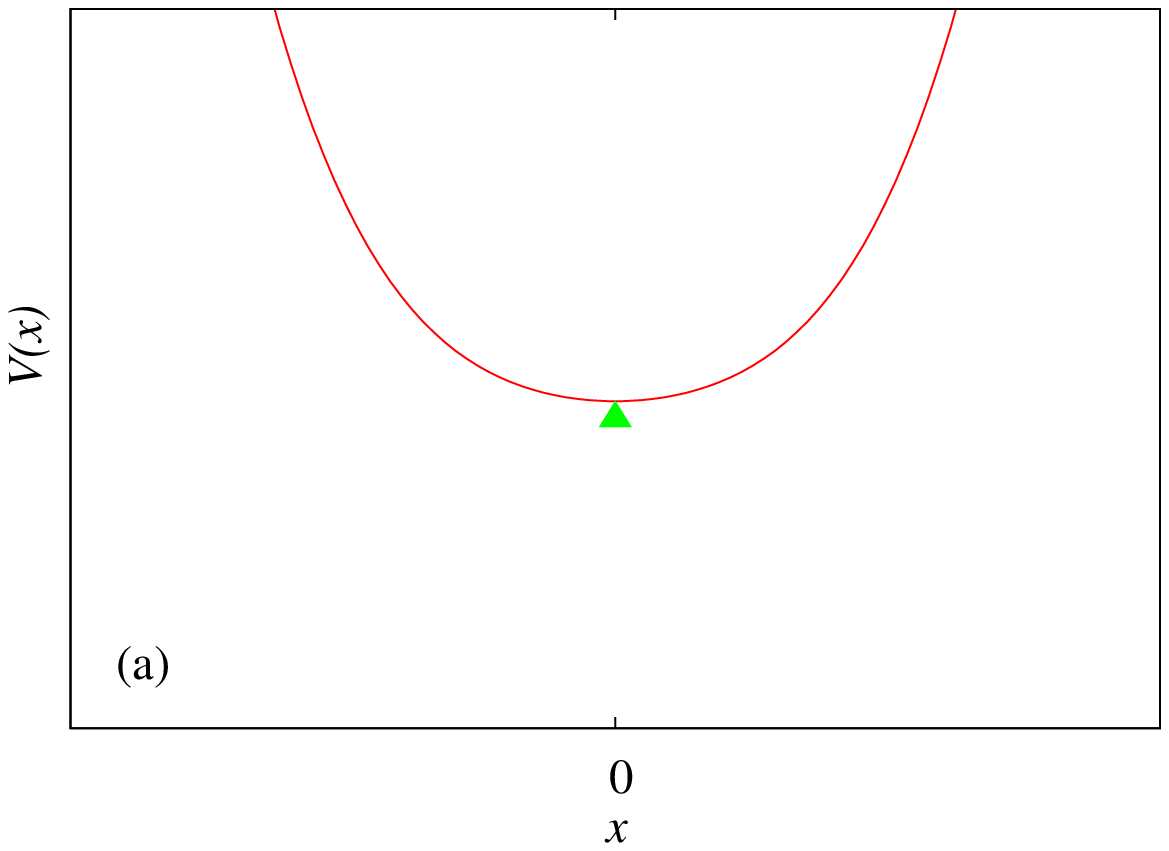}
\includegraphics[width=6.0cm]{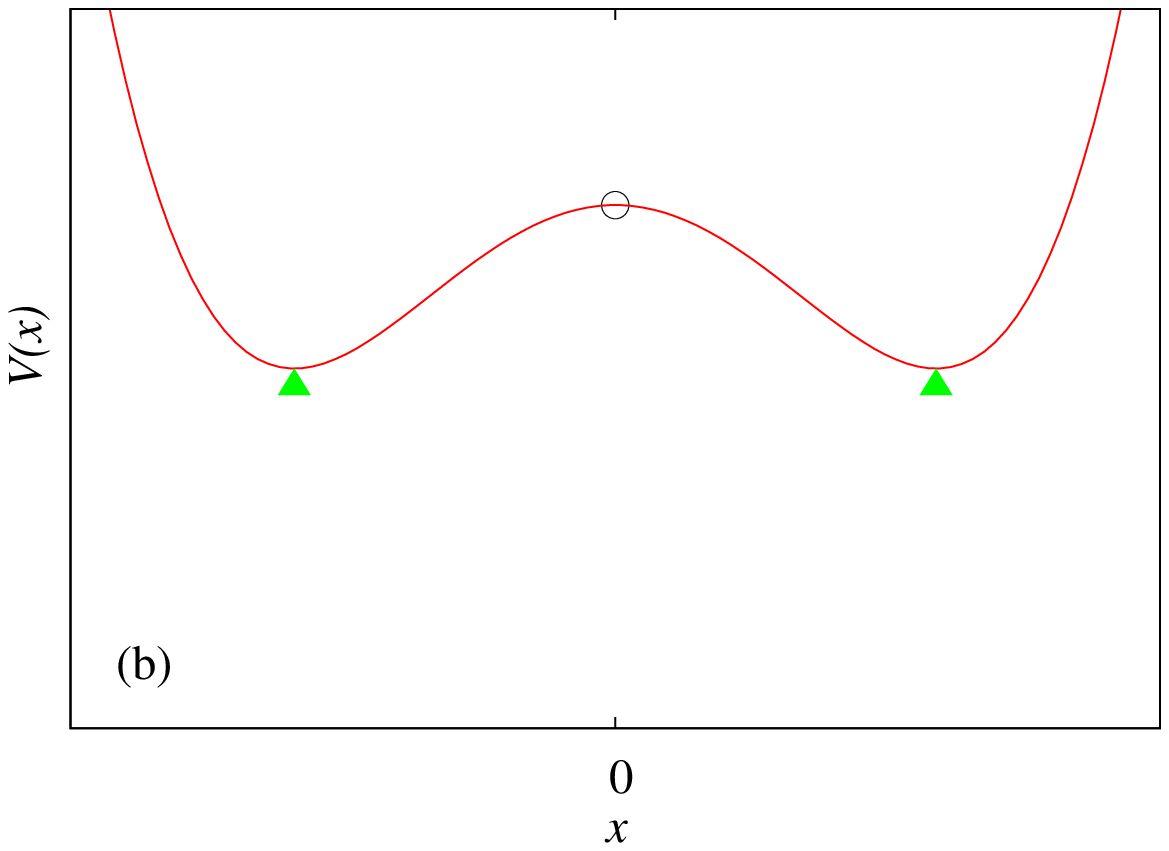}
\includegraphics[width=6.0cm]{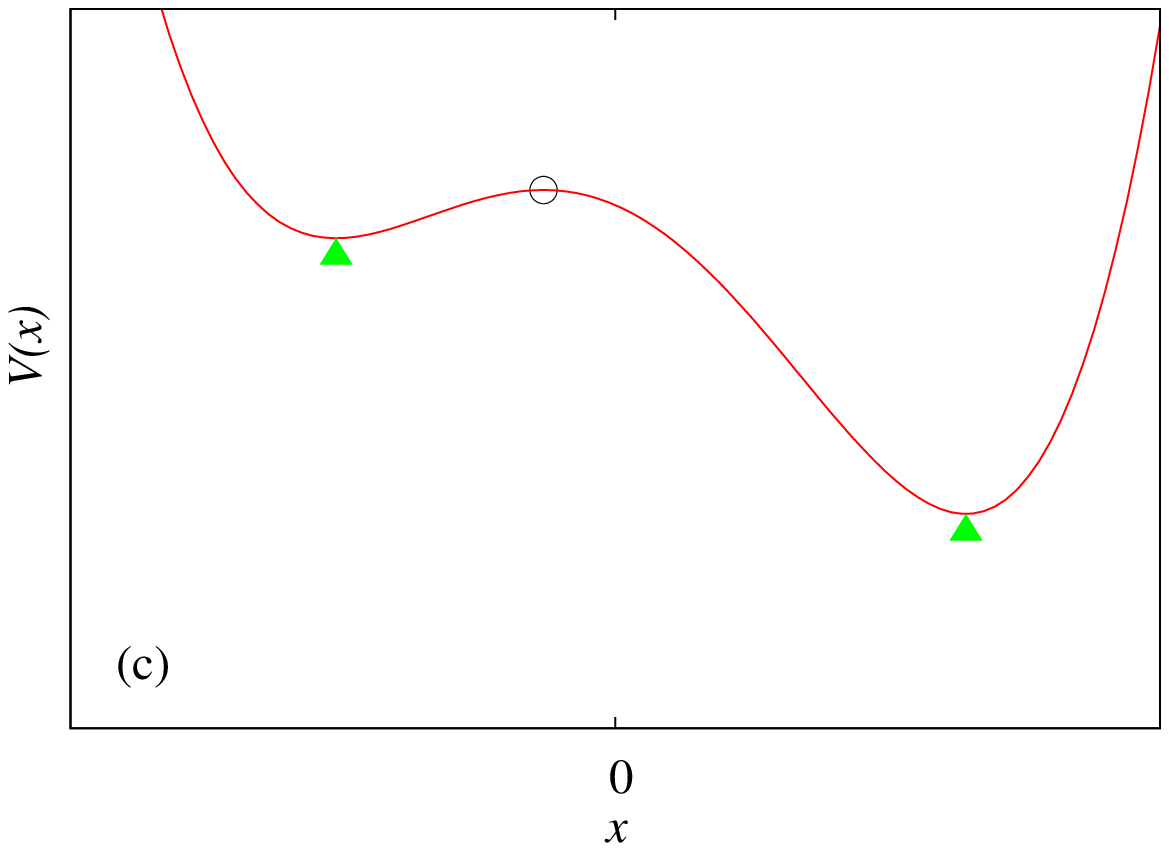}
\includegraphics[width=6.0cm]{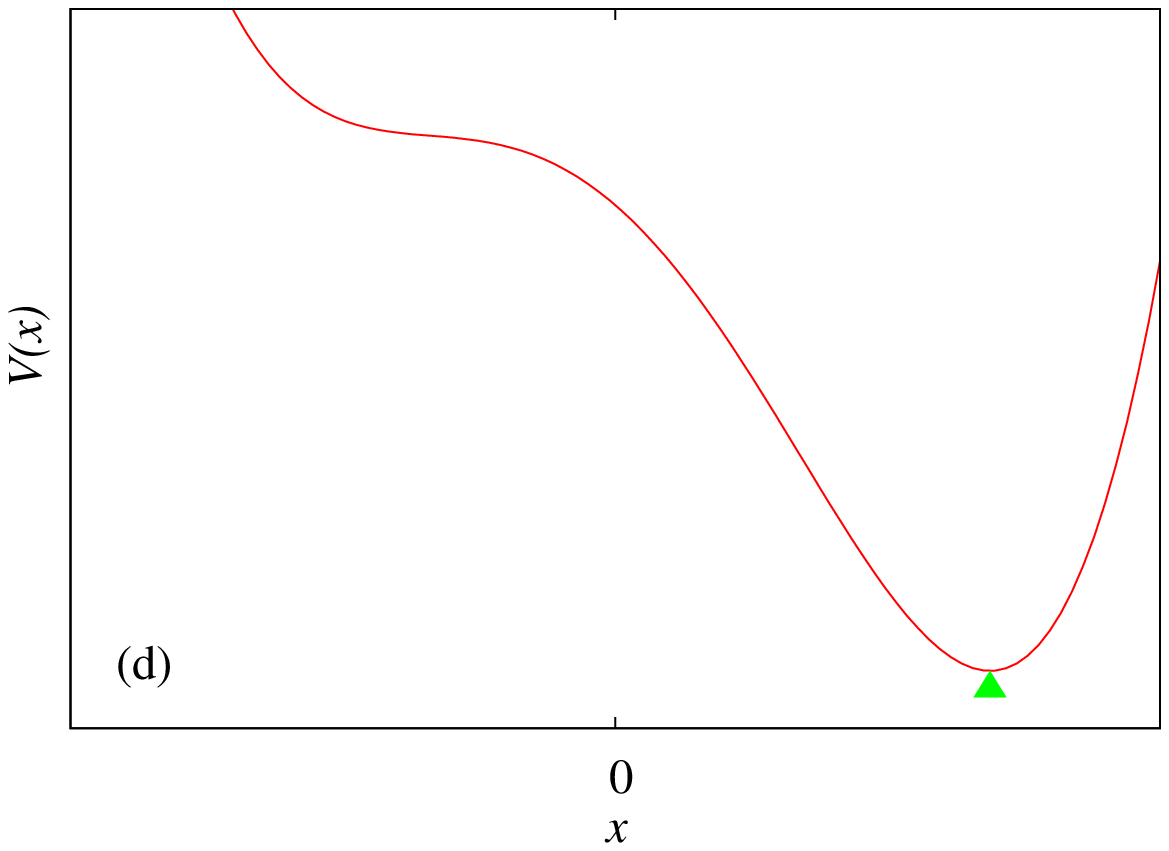}
\caption{Schematic diagram showing the stable points in the potential $V(x)= \varepsilon x + \eta x^2 + x^4$. The local minima of the potential are marked by triangles. The circles mark local maxima in the potential. Panel (a) shows the case of positive $\eta$, which gives a single local minimum. Here we have taken $\varepsilon=0$. Panels (b-d) show the potential for negative values of $\eta$ (specifically $\eta=-1$), which generally gives a double-well potential and hence two local minima of the potential. Panel (b) shows the symmetric case $\varepsilon=0$, where two equivalent local minima exist. Panel(c) shows the weakly asymmetric case $\varepsilon=-0.3$. In this case, two local minima exist. However, one of them corresponds to a metastable excited state. The curvature, and hence the excitation frequency, around the metastable excited state is smaller than that around the ground state. Panel (d) shows the strongly asymmetric case $\varepsilon=-0.6$, where the potential has only one minimum and there is no metastable excited state.}
\label{Fig:DoubleWellPotential}
\end{figure}

The number and nature of the solutions of Eqs.~(\ref{Eq:ClassicalGroundState}, \ref{Eq:ClassicalGroundStateX}) can be intuitively understood by considering the single-particle trapping potential of the form $V(x)= \varepsilon x + \eta x^2 + x^4$ \cite{DoubleWellPotentialJustification}. This potential, which is illustrated in Fig.~\ref{Fig:DoubleWellPotential}, exhibits bifurcation behavior. This behavior is most easily obtained by considering the symmetry point ($\varepsilon=0$): the potential has one or two minima depending on the sign of $\eta$. When $\eta$ is positive, there is only one minimum. When $\eta$ is negative, there are two local minima and they are equivalent to each other, because of the symmetry in $V(x)$. Between these two local minima, which correspond to dynamically stable solutions, there is a third solution of the equation $dV/dx=0$ (namely $x=0$) that is dynamically unstable. The bifurcation point $\eta=0$ corresponds to the critical point $4g^2N/(\hbar\omega\Delta)=1$. If we take the case of positive $\eta$ and move away from the symmetry point, the term $\varepsilon x$ causes the single minimum of the potential to be shifted away from $x=0$. When we move away from the symmetry point in the case of negative $\eta$, the term $\varepsilon x$ tilts the double-well potential, and one of the local minima becomes the global minimum, while the other local minimum becomes an excited metastable state. At a certain critical value of $\varepsilon$, the metastable excited solution disappears and the potential has a single minimum. The point where the metastable excited state disappears corresponds to the $\epsilon$ value that separates the regime where Eq.~(\ref{Eq:ClassicalGroundStateX}) has one solution and the regime where Eq.~(\ref{Eq:ClassicalGroundStateX}) has three solutions.

We finally note that in the regime $4g^2N/(\hbar\omega\Delta)>1$ with small $|\epsilon|$, where multiple dynamically stable states exist, it is natural to say that the system is in the superradiant phase with nontrivial qubit-cavity correlations. When $|\epsilon|$ is large such that there is only one stable state, it becomes less meaningful to say that the system is still in the superradiant phase. According to the definitions used in this work, we say that the system goes back to the normal phase when $|\epsilon|$ exceeds a certain ($g$-dependent) critical value.

\section{Spectrum in the absence of fluctuations}
\label{Sec:Spectrum}

If a probe signal is applied to the system in its ground state, it will exhibit some spectral response at the frequencies that correspond to the excitation modes of the system. Since in the case of identical qubits we effectively have two degrees freedom, we expect to have two excitation modes. These can be found using the Holstein-Primakoff transformation, where we replace the spin operators by harmonic oscillator operators \cite{HolsteinPrimakoff}. One possibility is to use the operators $\hat{S}_x$ and $\hat{S}_z$ in applying the transformation, and the ensuing derivations would closely follow Refs.~\cite{EmaryPRE,EmaryPRA}. We follow an alternative approach that gives the same results.

We start by rotating the reference frame for the spin operators:
\begin{eqnarray}
\tilde{S}_x & = & \cos\theta \hat{S}_x - \sin\theta \hat{S}_z
\nonumber \\
\tilde{S}_z & = & \cos\theta \hat{S}_z + \sin\theta \hat{S}_x,
\end{eqnarray}
and making the transformation $\tilde{x}=\hat{x}-x_0$, $\tilde{p}_x=\hat{p}_x$. Here $x_0$ is the ground-state value of $x$ obtained from Eq.~(\ref{Eq:ClassicalGroundStateX}), and $\theta$ is the corresponding value of $\theta$ obtained from Eq.~(\ref{Eq:ClassicalGroundState}). Upon making these transformations, the Hamiltonian is transformed into the form
\begin{widetext}
\begin{eqnarray}
\hat{H} & = & \Delta \left( \cos\theta \tilde{S}_z - \sin\theta \tilde{S}_x \right) + \epsilon \left( \cos\theta \tilde{S}_x + \sin\theta \tilde{S}_z \right) + \hbar\omega \left( \tilde{x} + x_0 \right)^2 + \hbar\omega \tilde{p}_x^2 + 4 g \left( \cos\theta \tilde{S}_x + \sin\theta \tilde{S}_z \right) \left( \tilde{x} + x_0 \right)
\nonumber \\
& = & \Big( \Delta \cos\theta + (\epsilon + 4 g x_0) \sin\theta \Big) \tilde{S}_z + \hbar\omega \tilde{x}^2 + \hbar\omega \tilde{p}_x^2 + 4 g \cos\theta \tilde{S}_x \tilde{x} + \hbar\omega x_0^2 \nonumber \\ & & + \Big( (\epsilon + 4 g x_0) \cos\theta - \Delta \sin\theta \Big) \tilde{S}_x + \left( 4 g \sin\theta \tilde{S}_z + 2 \hbar\omega x_0 \right) \tilde{x}.
\label{Eq:HamiltonianAroundLocalMimimum}
\end{eqnarray}
\end{widetext}
We know that if $\theta$ and $x_0$ satisfy Eq.~(\ref{Eq:ClassicalGroundState}), the coefficients of $\tilde{S}_x$ and $\tilde{x}$ in the last two terms in the last line of Eq.~(\ref{Eq:HamiltonianAroundLocalMimimum}) vanish to first order in $\tilde{S}_x$ and $\tilde{x}$. Furthermore, $\tilde{S}_x=\tilde{x}=0$ in the classical ground state. As a result, the first non-vanishing term in the last two terms in Eq.~(\ref{Eq:HamiltonianAroundLocalMimimum}) appears at third order in $\tilde{S}_x$ and/or $\tilde{x}$. The term before them is a constant. We can therefore ignore these terms and consider the Hamiltonian
\begin{eqnarray}
\hat{H} & = & \left( \Delta \cos\theta + (\epsilon + 4 g x_0) \sin\theta \right) \tilde{S}_z \nonumber \\ & & + \hbar\omega \tilde{x}^2 + \hbar\omega \tilde{p}_x^2 + 4 g \cos\theta \tilde{S}_x \tilde{x}.
\end{eqnarray}
Although so far in this section we have said that we are considering the classical ground state, the derivation remains valid when considering the metastable excited state in the bistability regime.

Now we perform the Holstein-Primakoff transformation:
\begin{eqnarray}
\tilde{S}_z & = & \hat{b}^{\dagger} \hat{b} - \frac{N}{2}
\nonumber \\
\tilde{S}_x & = & \frac{1}{2} \left( \tilde{S}_+ + \tilde{S}_- \right)
\nonumber \\
\tilde{S}_+ & = & \hat{b}^{\dagger} \sqrt{N - \hat{b}^{\dagger} \hat{b}}
\nonumber \\
\tilde{S}_- & = & \tilde{S}_+^{\dagger},
\end{eqnarray}
where $\hat{b}$ and $\hat{b}^{\dagger}$ are harmonic oscillator operators. It is useful here to define the operators
\begin{equation}
\tilde{y} = \frac{ \hat{b} + \hat{b}^\dagger }{2}
\end{equation}
and
\begin{equation}
\tilde{p}_y = \frac{ \hat{b} - \hat{b}^\dagger }{2i}.
\end{equation}
The Hamiltonian can now be expressed as
\begin{widetext}
\begin{eqnarray}
\hat{H} & = & \left( \Delta \cos\theta + (\epsilon + 4 g x_0) \sin\theta \right) \left( \hat{b}^{\dagger} \hat{b} - \frac{N}{2} \right) + \hbar\omega \left( \tilde{x}^2 + \tilde{p}_x^2 \right) + 2 g \cos\theta \left( \hat{b}^{\dagger} \sqrt{N - \hat{b}^{\dagger} \hat{b}} + \sqrt{N - \hat{b}^{\dagger} \hat{b}} \hat{b} \right) \tilde{x}
\nonumber \\
& \approx & \left( \Delta \cos\theta + (\epsilon + 4 g x_0) \sin\theta \right) \left( \tilde{y}^2 + \tilde{p}_y^2 \right) + \hbar\omega \left( \tilde{x}^2 + \tilde{p}_x^2 \right) + 4 g \sqrt{N} \cos\theta \tilde{x} \tilde{y} + {\rm constant}.
\label{Eq:TwoModeHamiltonian}
\end{eqnarray}
\end{widetext}
This Hamiltonian describes two bi-linearly coupled harmonic oscillators. One can redefine the variables in such a Hamiltonian such that the coupling term is eliminated and the Hamiltonian describes two uncoupled harmonic oscillators. These are the normal oscillation modes of the system. The eigen-energies of the system then have the form $n_+\hbar\nu_+ + n_-\hbar\nu_- + {\rm constant}$, where $n_{\pm}$ are two non-negative integers and $\nu_{\pm}$ are the frequencies of the oscillation modes, just as one would expect for the combined energy eigenstates of two uncoupled harmonic oscillators. Note that Eq.~(\ref{Eq:TwoModeHamiltonian}) is valid in both the normal and superradiant phases. The difference between the normal and superradiant cases enters in the coefficients of the different operators in Eq.~(\ref{Eq:TwoModeHamiltonian}). For example, taking $\epsilon=0$, in the normal phase, $\theta=0$ and $x_0=0$, while in the superradiant phase, $x_0$ is given by Eq.~(\ref{Eq:CavityFieldInSuperradiantState}) and $\theta$ is given by Eq.~(\ref{Eq:ClassicalGroundState}), which gives $\cos\theta=\hbar\omega\Delta/(4g^2N)$. Furthermore, in the superradiant case, there can be two sets of parameters that correspond to stable or metastable states that will contribute spectral lines observed in experiment. For each set of parameters entering in Eq.~(\ref{Eq:TwoModeHamiltonian}), the oscillation frequencies of the two oscillation modes are given by
\begin{equation}
\nu_{\pm}^2 = \frac{1}{2} \left( \gamma^2 + \omega^2 \pm \sqrt{(\gamma^2 - \omega^2)^2 + 16 g^2 N \gamma\omega \cos^2 \theta/\hbar^2} \right)
\label{Eq:OscillationModeFrequencies}
\end{equation}
where we have defined $\gamma=(\Delta \cos\theta + (\epsilon + 4 g x_0) \sin\theta)/\hbar$. The frequency $\nu_+$ corresponds to the high-frequency mode while $\nu_-$ corresponds to the low-frequency mode in the spectrum.

We now use Eq.~(\ref{Eq:OscillationModeFrequencies}), with $x_0$ calculated from Eq.~(\ref{Eq:ClassicalGroundStateX}), to calculate the frequencies of the oscillation modes for a number of different parameter combinations. In some cases Eq.~(\ref{Eq:ClassicalGroundStateX}) has only one solution and we obtain only two resonance frequencies. In other cases, specifically in the superradiant phase with small values of $|\epsilon|$, we find two stable solutions of Eq.~(\ref{Eq:ClassicalGroundStateX}), and we therefore obtain four resonance frequencies.

\begin{figure}[h]
\includegraphics[width=6.0cm]{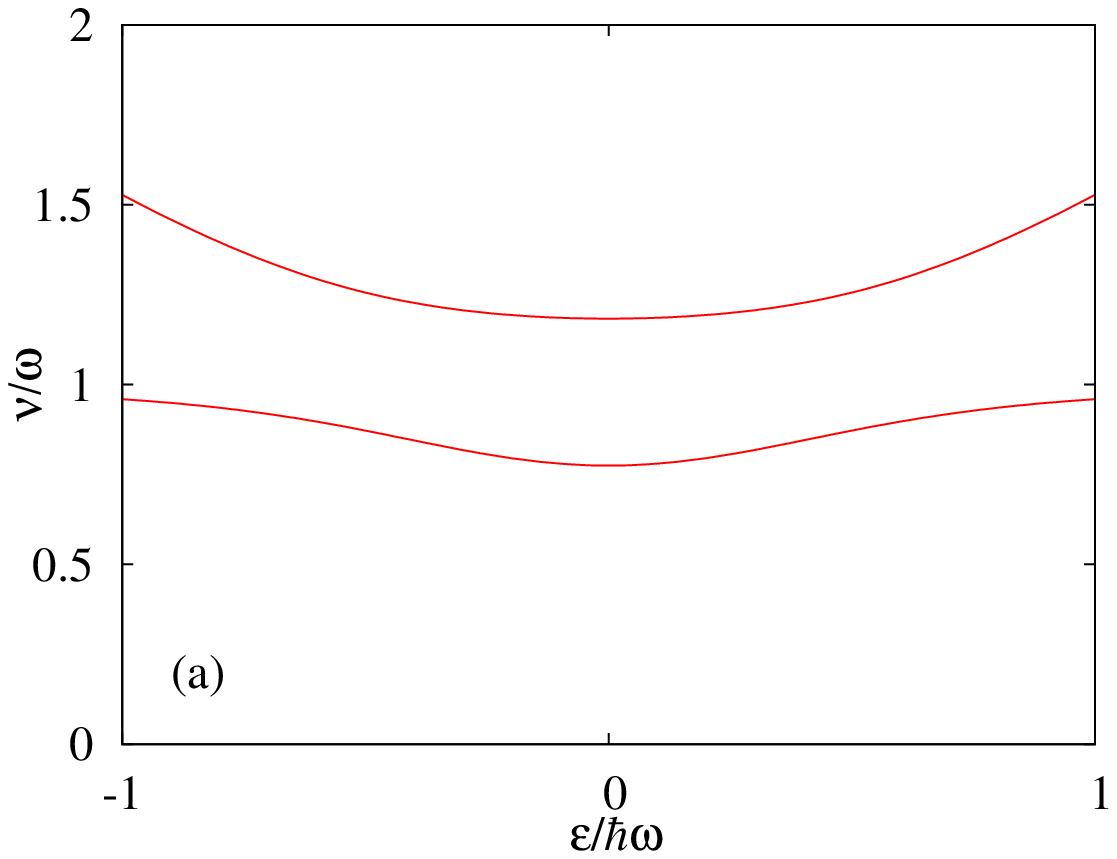}
\includegraphics[width=6.0cm]{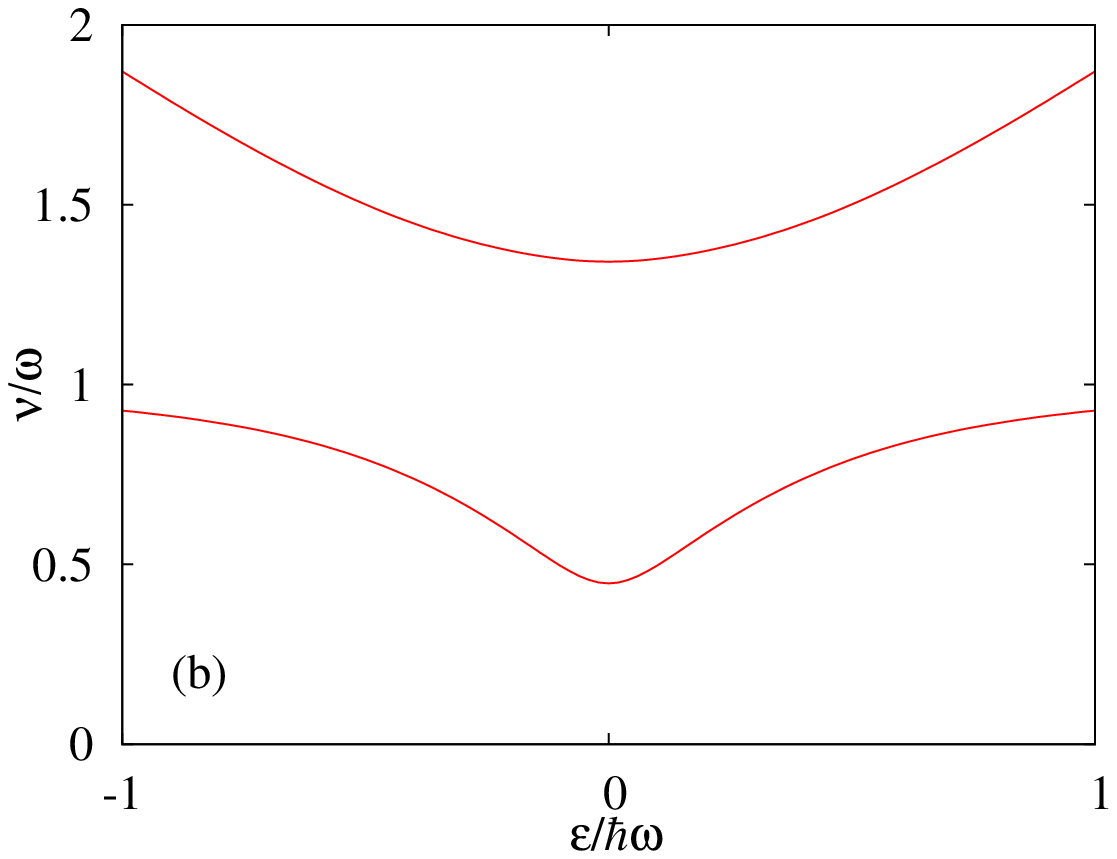}
\includegraphics[width=6.0cm]{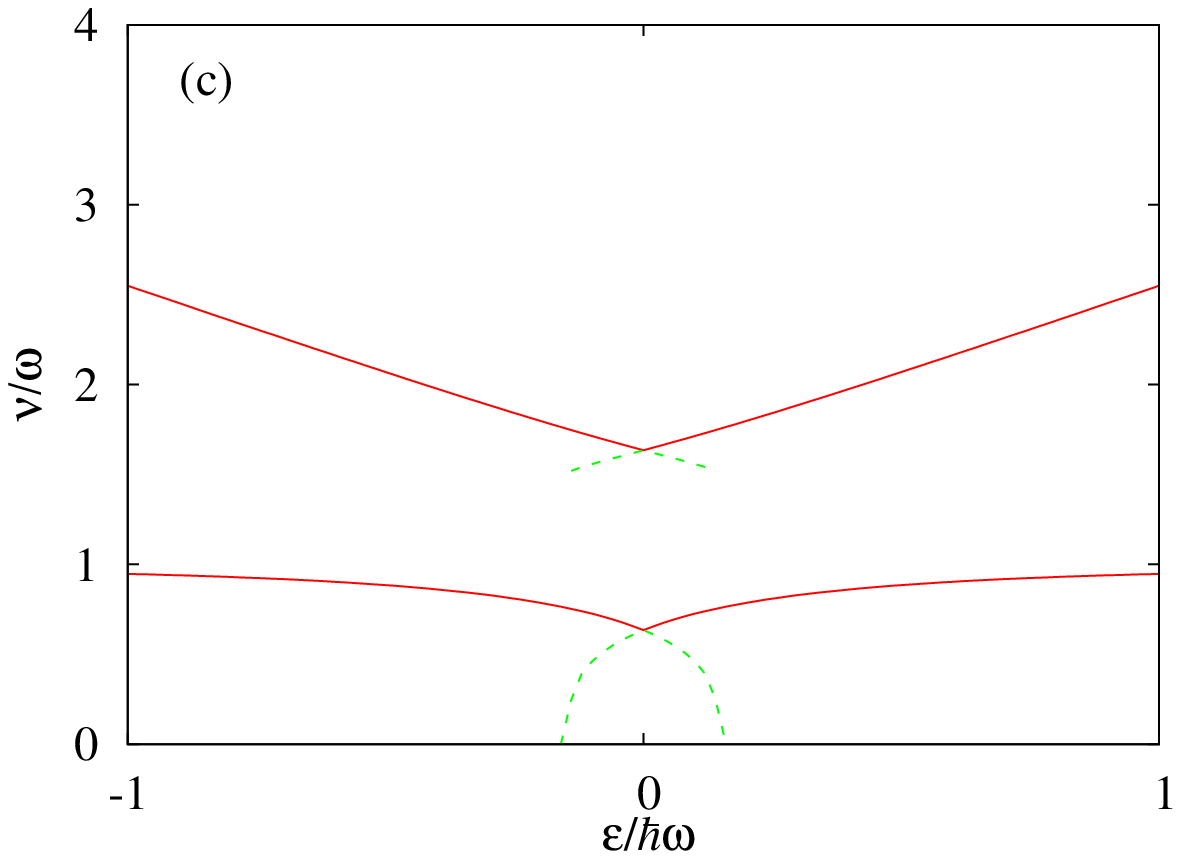}
\includegraphics[width=6.0cm]{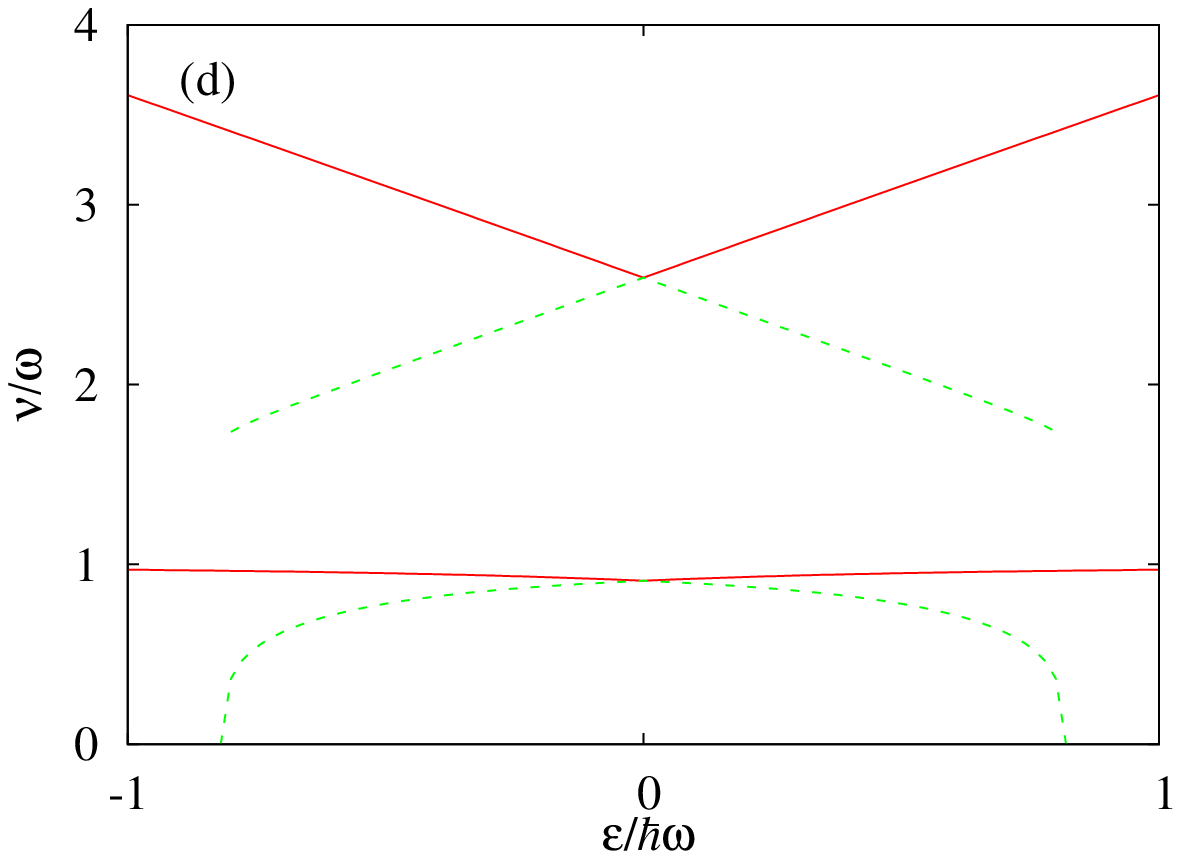}
\caption{Frequencies of the excitation modes $\nu_{\pm}$ as functions of the bias parameter $\epsilon/(\hbar\omega)$. The red solid lines are the spectral lines that correspond to excitation from the ground state, while the green dashed lines correspond to excitation from the metastable excited state. Here we set $\Delta=\hbar\omega$. The coupling strength is given by $\sqrt{N}g/(\hbar\omega)=0.2$ (a), $0.4$ (b), 0.6 (c) and 0.8 (d). In (a) and (b), $\sqrt{N}g/(\hbar\omega)<0.5$, which corresponds to the normal phase, while in (c) and (d) $\sqrt{N}g/(\hbar\omega)>0.5$, which correspond to the superradiant phase. In the normal phase only two spectral lines are obtained, whereas in the superradiant phase the existence of a metastable excited state results in two additional spectral lines. As we move away from the symmetry point, there is a critical value of $|\epsilon|$ at which the metastable excited state does not exist any more, and its spectral lines disappear. In particular, we note that the high-frequency dashed line disappears at the same value of $\epsilon$ where the frequency of the lowest spectral line reaches zero, even though this fact might not be entirely clear from the appearance of the lines in the figure.}
\label{Fig:ResonantCase}
\end{figure}

In Fig.~\ref{Fig:ResonantCase} we plot the resonance frequencies as functions of $\epsilon$ for the normal and superradiant phases when $\Delta=\hbar\omega$. In the normal phase (Fig.~\ref{Fig:ResonantCase}(a,b)), we obtain the usual circuit-QED spectrum: away from the symmetry point the oscillator's resonance frequency is $\omega$, while the qubit excitation frequency is approximately $\epsilon/\hbar$. The two lines deviate from this simple behavior in the vicinity of the symmetry point because of the hybridization of the oscillator and qubit excitation modes. For example, the spectral line that coincides with the resonator's frequency at $\epsilon\rightarrow\infty$ has a dip at the symmetry point, which is the point where the qubit's resonance frequency approaches it from above. The spectrum in the superradiant phase is qualitatively different. In Fig.~\ref{Fig:ResonantCase}(c,d) we include the resonance frequencies for the solutions of Eq.~(\ref{Eq:ClassicalGroundStateX}) that correspond to both the ground state and the metastable excited state. The ground state has two excitation modes as in Fig.~\ref{Fig:ResonantCase}(a,b). The metastable state also has two excitation modes with frequencies that are different from those of the ground state, except at the symmetry point where the two states become degenerate and hence equivalent ground states. As a result, there is now a range of $\epsilon$ values where we have four excitation mode frequencies. The two pairs of excitation frequencies in the spectrum exhibit crossings at the symmetry point.

As in Sec.~\ref{Sec:Model}, the behavior of the two low-frequency modes in the superradiant phase can be intuitively understood by thinking of the ground state and metastable excited state in the double-well potential $V(x)= \varepsilon x - x^2 + x^4$ (Fig.~\ref{Fig:DoubleWellPotential}). At the symmetry point ($\varepsilon=0$), the two local minima are equivalent to each other and the curvature of the potential $V(x)$ is the same at these two points. If we move away from the symmetry point, the term $\varepsilon x$ tilts the potential, one of the local minima becomes deeper and its curvature increases, while the other local minimum becomes shallower. Because of the asymmetry created between the ground state and the metastable excited state, each of the two spectral lines is split into two, resulting in four spectral lines in total. As the tilting of the potential keeps increasing, the local minimum corresponding to the metastable excited state keeps becoming shallower until at some tilting slope it disappears and the potential has a single minimum. The disappearance of the metastable excited state coincides with the disappearance of two spectral lines from the spectrum, which occurs at large values of $|\epsilon|$. The point where the metastable excited state disappears also corresponds to the $\epsilon$ value that separates the regime where Eq.~(\ref{Eq:ClassicalGroundStateX}) has one solution and the regime where Eq.~(\ref{Eq:ClassicalGroundStateX}) has three solutions. As the tilting strength approaches the critical point, the shallow well that traps the metastable solution becomes increasingly shallow with the characteristic oscillation frequency reaching the value zero at the critical point. In practice, the occupation probability of the metastable excited state will decrease and vanish before we reach that point (because of either thermal or quantum fluctuations). As a result, the spectral line will fade and disappear before it reaches zero frequency. If we assume that even a small value of $|\epsilon|$ will eliminate the population of the excited state in a specific experimental setup, the dashed lines in Fig.~\ref{Fig:ResonantCase}(c,d) disappear and the spectrum exhibits two V-shaped spectral lines (corresponding to the two oscillation modes of the ground state) with sharp cusps at the symmetry point. Such V-shaped spectral lines were obtained in Ref.~\cite{EmaryPRA}, which did not consider the metastable excited state.

As mentioned above, the high frequency mode is mostly a qubit excitation mode. One might notice that in Fig.~\ref{Fig:ResonantCase}(c,d) even at the symmetry point the frequency of this mode is now well above the frequency of the other mode. The reason is that in the superradiant phase $x_0$ has a finite value even when the externally applied bias corresponds to the symmetry point, and as a result each qubit is effectively biased away from the symmetry point, leading to an increased resonance frequency. The frequency of this oscillation mode does not vanish even at the critical value of $\epsilon$ at which the low frequency reaches zero. However, since there is no metastable state beyond that point, these two spectral lines will disappear together. If the probe field couples to the oscillator, one will mainly observe the response from the low-frequency mode, because that is the mode that has mostly cavity-excitation character while the high-frequency mode has a mostly qubit-excitation character.

\begin{figure}[h]
\includegraphics[width=8.0cm]{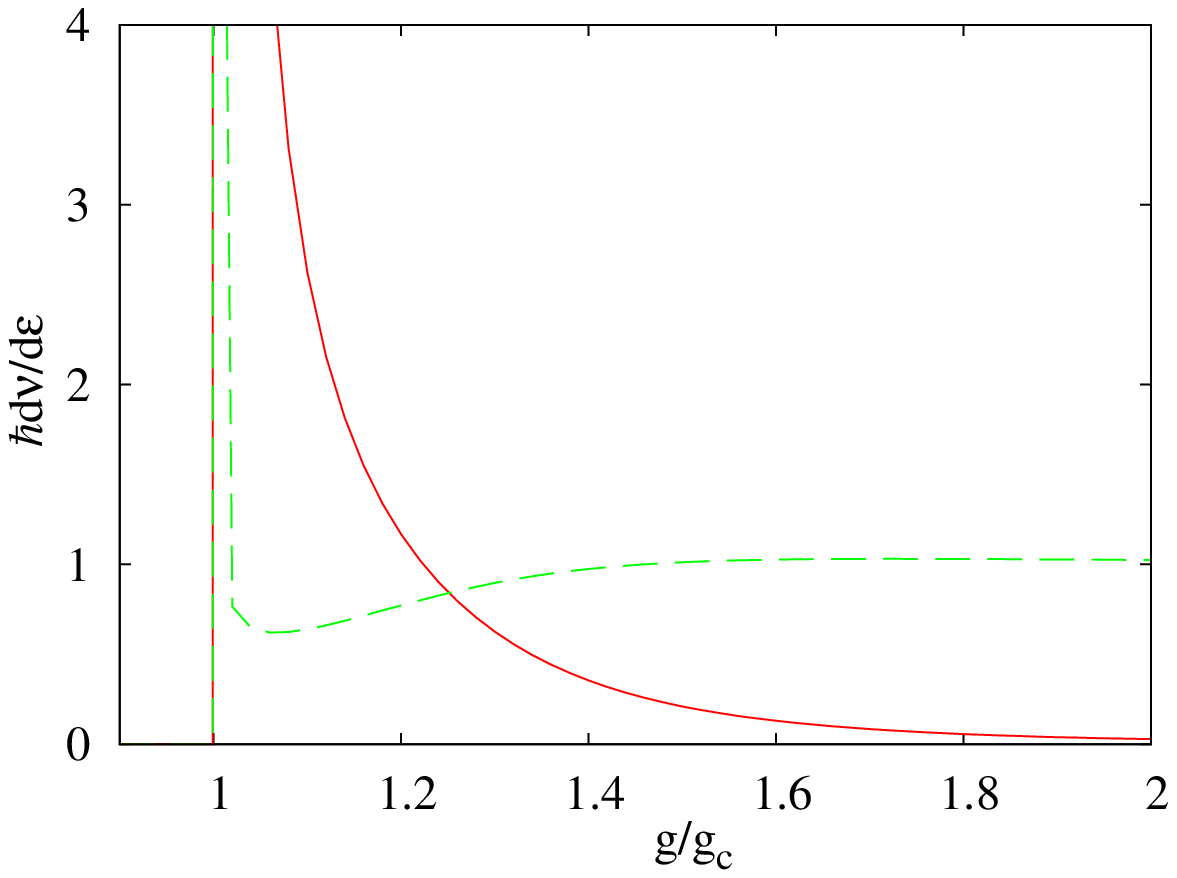}
\caption{Slopes of the frequencies $\nu_-$ (red solid line) and $\nu_+$ (green dashed line) with respect to $\epsilon$ (i.e.~$\hbar d\nu_{\pm}/d\epsilon$) at the symmetry point as functions of coupling strength $g$ (measured relative to the critical value $g_c$). We take the slopes of the spectral lines that correspond to the ground state for positive $\epsilon$, i.e.~those that extend to $\epsilon\rightarrow\infty$. As in Fig.~\ref{Fig:ResonantCase} we set $\Delta=\hbar\omega$. In the normal phase (i.e.~when $g/g_c<1$), the slopes vanish, because the spectral lines have their minima at the symmetry point. In the superradiant phase, two pairs of spectral lines cross with slopes that are equal in magnitude but with opposite signs. As we approach the transition point $g=g_c=\sqrt{\Delta\hbar\omega/(4N)}$ from above, the slope of $\nu_-$ diverges as $(g-g_c)^{-1}$.}
\label{Fig:SlopeAtSymmetryPointResonant}
\end{figure}

\begin{figure}[h]
\includegraphics[width=8.0cm]{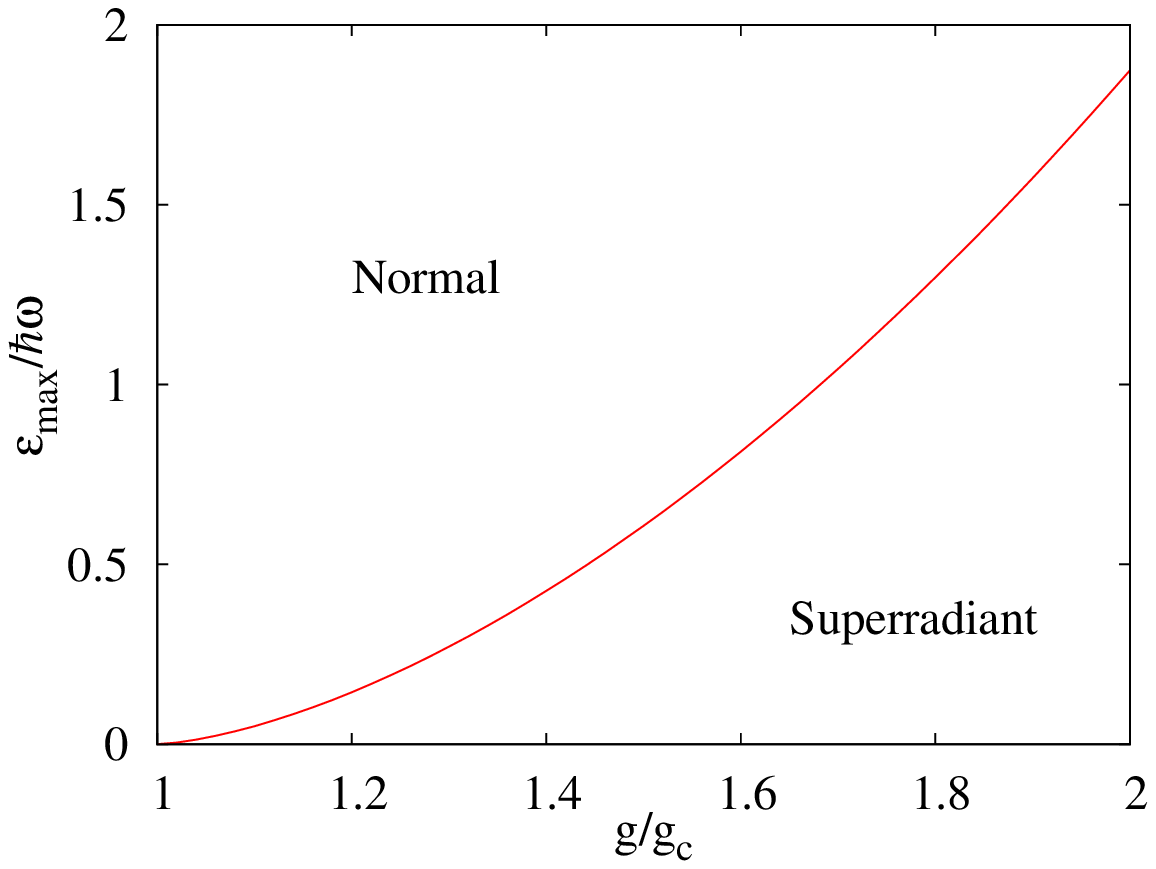}
\caption{The maximum value of $|\epsilon|$ at which there exists a metastable excited state, i.e.~the maximum value of $|\epsilon|$ at which one expects to see four lines in the spectrum, as a function of $g/g_c$. As above $\Delta=\hbar\omega$. Alternatively, this figure can be seen as a phase diagram: the labels ``Normal" and ``Superradiant" describe the phase that is obtained in each region in the $g$-$\epsilon$ parameter space, where the superradiance region is defined by the existence of two stable states.}
\label{Fig:BistabilityRangeResonant}
\end{figure}

Since the frequency range that is accessible with typical measurement methods is often limited (in the range of a few GHz for superconducting qubit circuits), it can be difficult or impossible to observe the frequency of the oscillator mode as it goes all the way down close to zero. Even in this case, the observation of V- or X-shaped spectral lines (at and around the symmetry point), as opposed to the smooth dip obtained in the normal phase, can serve as a signature of the superradiant phase. For the experimental observation of the V or X shape, it is crucial to have sufficiently large slopes of the spectral lines at the symmetry point. In Fig.~\ref{Fig:SlopeAtSymmetryPointResonant} we plot the slopes of both the low- and high-frequency spectral lines at the symmetry point. Note that each pair of intersecting spectral lines have the same slope but with opposite signs at the symmetry point, and we shall therefore refer to a single slope for each pair, specifically taking the spectral line that extends to $\epsilon\rightarrow\infty$. As we can see from the figure, just above the critical point, the slopes of both lines are very large, which is desirable for experimental purposes. The slope of the $\nu_+$ line quickly drops and converges to the value one, while the slope of the $\nu_-$ line decreases significantly more slowly, asymptotically approaching zero. It is in fact not very surprising that the slope of the $\nu_-$ line diverges as $g\rightarrow g_c^+$, because as can be seen from Fig.~\ref{Fig:ResonantCase} the spectral line frequencies go down to zero with infinite slope in the superradiant phase, and at $g\rightarrow g_c^+$ the edges of the two spectral lines (i.e.~$\nu_-$ for the two stable states) drop to zero at $\epsilon=0$. The fact that the frequency $\nu_-$ is very small just above the critical point is undesirable for purposes of experimental observation, because this small frequency might be outside the measurable frequency range, as mentioned above. One could use more advanced spectroscopy techniques to measure these low frequencies, such as the two-tone spectroscopy used recently in Ref.~\cite{YoshiharaLamb}. On the other hand, for $g$ values that are much higher than the critical value for superradiance, the slope of the two $\nu_-$ lines asymptotically approaches zero, which makes the two spectral lines difficult to resolve from one another. As a result, if one is probing the low-frequency spectral line, it can be better not to have parameters that are very deep in the superradiant phase, where the spectral response goes back to resembling the spectrum of an isolated harmonic oscillator, as discussed in Ref.~\cite{YoshiharaSpectra}. The slope of the $\nu_+$ lines has a weak dependence on $g$ after the sharp peak just above $g_c$. One could therefore think of this slope as making a sudden jump from zero (in the normal phase) to one (in the superradiant phase). The $\nu_+$ spectral lines can therefore clearly distinguish between the normal and superradiant phases, if these lines can be observed in a specific experiment setup.

Another possibly important parameter for purposes of experimental observation is the range of $\epsilon$ values for which the four spectral lines exist. This quantity is plotted in Fig.~\ref{Fig:BistabilityRangeResonant}. The maximum separation between the two spectral lines in the upper branch, which occurs at the maximum value of $|\epsilon|$ before one of the two lines disappears (i.e.~the value of $|\epsilon|$ plotted in Fig.~\ref{Fig:BistabilityRangeResonant}) is shown in the Appendix.

\begin{figure}[h]
\includegraphics[width=6.0cm]{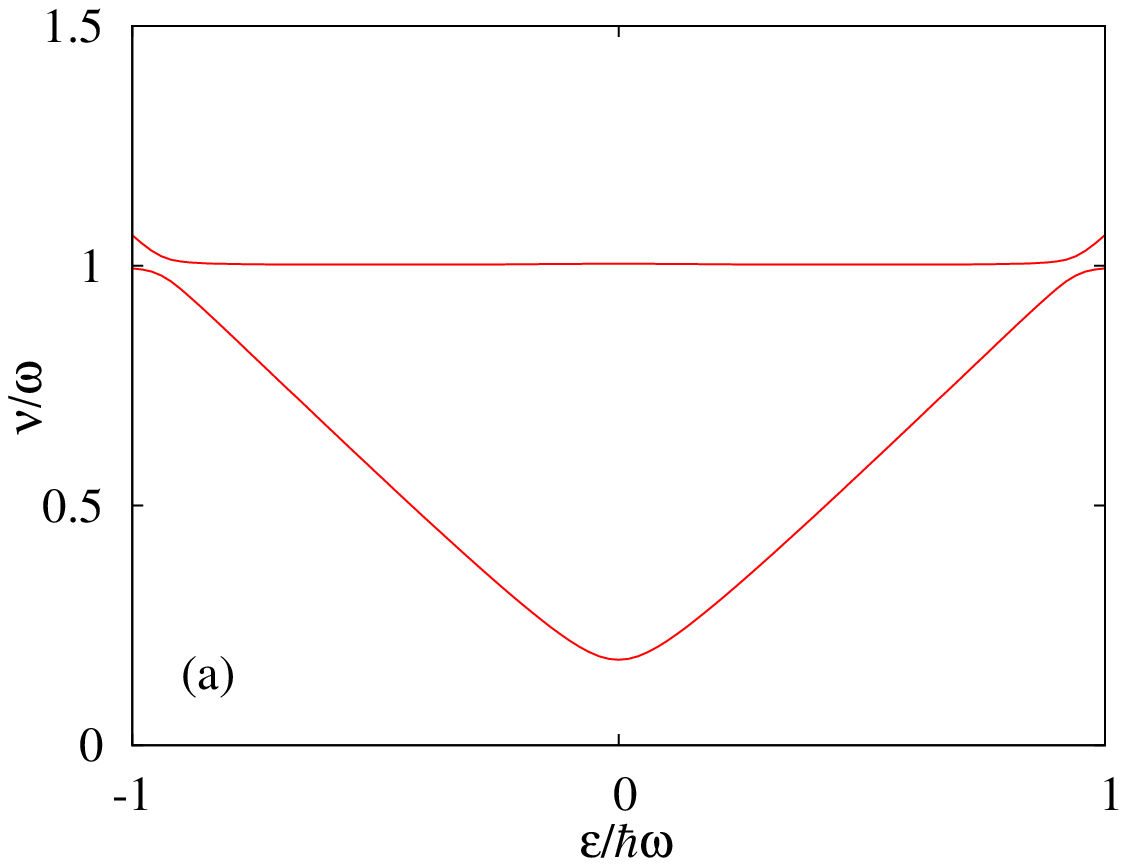}
\includegraphics[width=6.0cm]{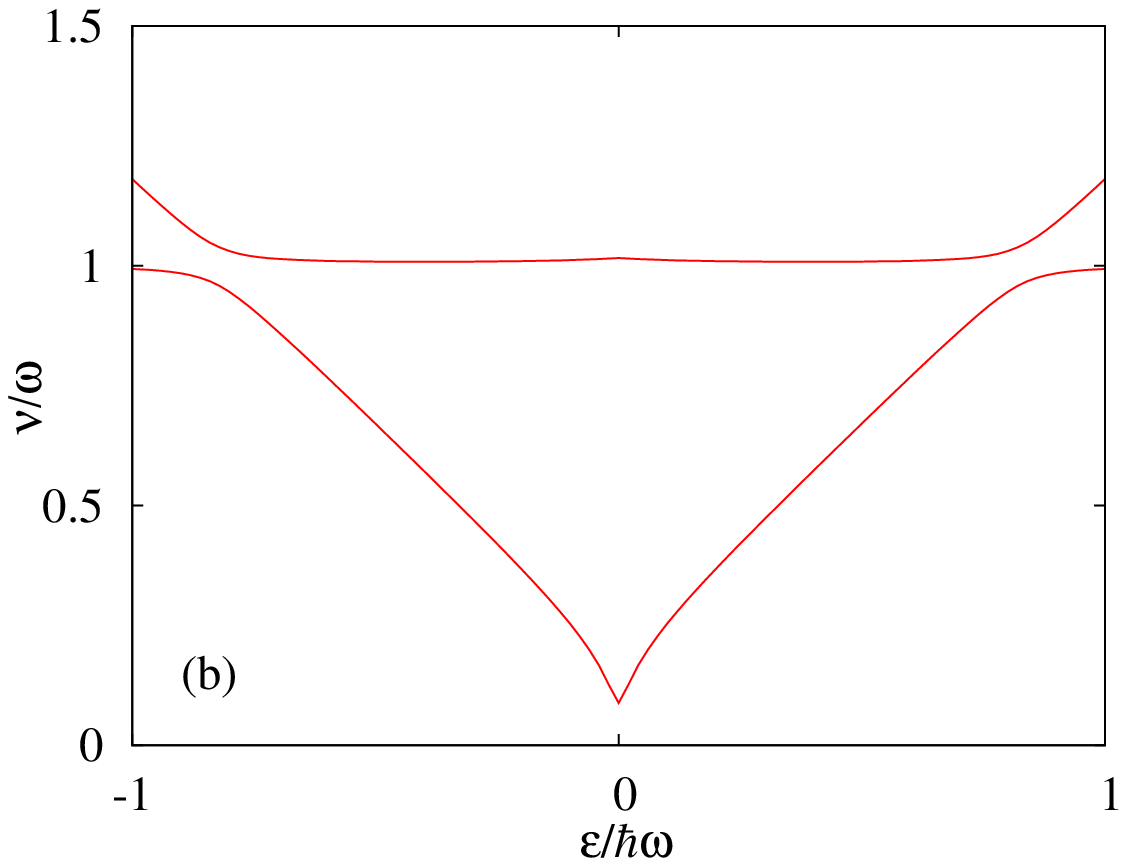}
\includegraphics[width=6.0cm]{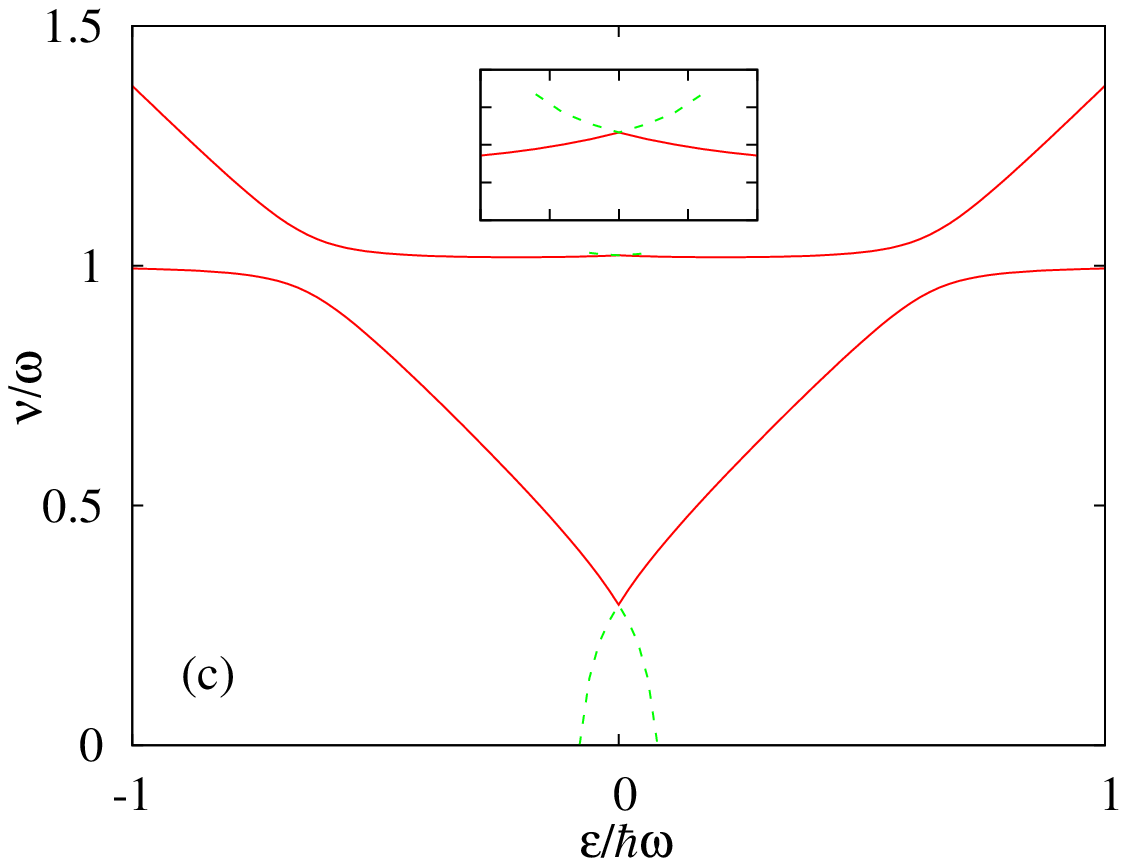}
\includegraphics[width=6.0cm]{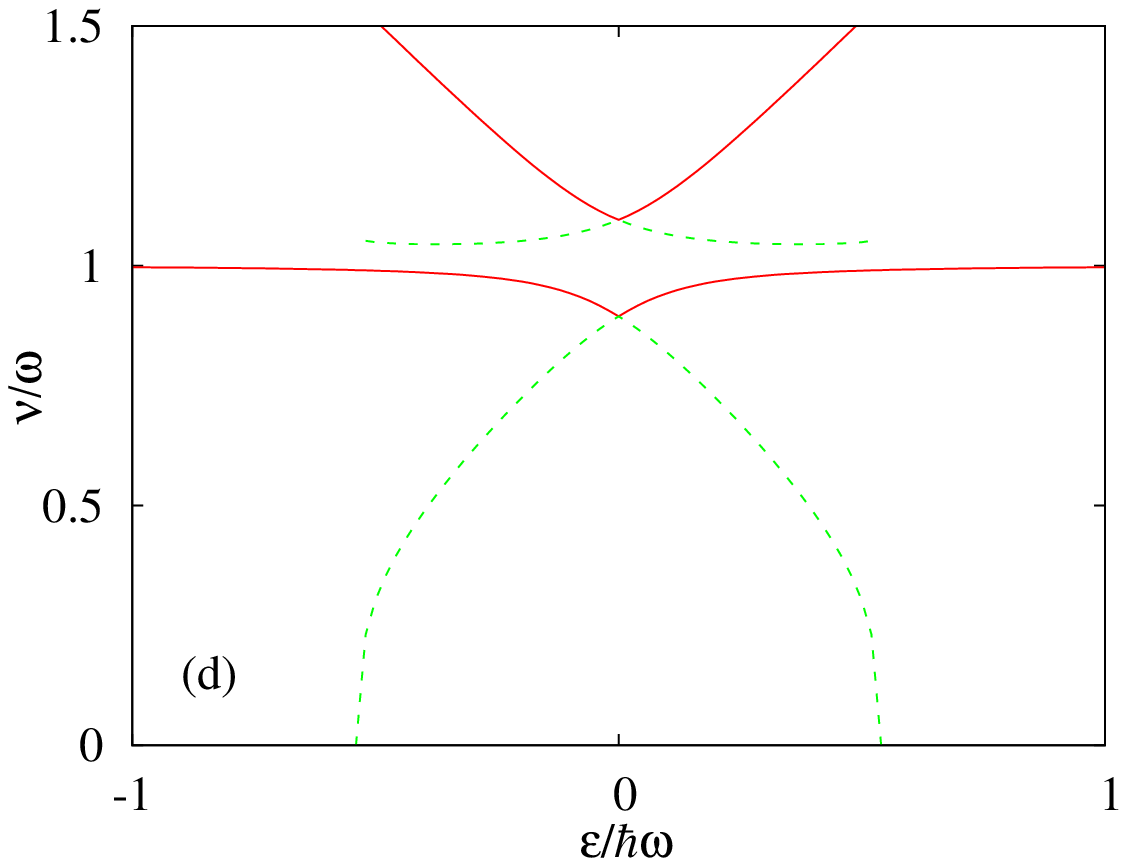}
\caption{Frequencies of the excitation modes as functions of $\epsilon/(\hbar\omega)$. As in Fig.~\ref{Fig:ResonantCase}, the red solid lines are the spectral lines that correspond to excitation from the ground state, while the green dashed lines correspond to excitation from the metastable excited state. Here we take the case of small $\Delta$. Specifically we set $\Delta/\hbar\omega=0.2$, which gives $g_c/(\sqrt{N}\hbar\omega)=0.224$. We set $\sqrt{N}g/(\hbar\omega)=0.1$ (a), 0.2 (b), 0.3 (c) and 0.5 (d). The inset in (c) shows a magnified view of the region $x\in [-0.1,0.1]$, $y\in [1.01,1.03]$.}
\label{Fig:SmallDeltaCase}
\end{figure}

So far we have focused on the special case $\Delta=\hbar\omega$. We now take the case where $\Delta$ is much smaller than $\hbar\omega$ (Fig.~\ref{Fig:SmallDeltaCase}), as in the experiment of Ref.~\cite{Kakuyanagi}. Three features are worth noting here. (1) when we first enter the superradiant phase (Fig.~\ref{Fig:SmallDeltaCase}(c)), the spectral line of the high oscillation frequency for the metastable excited state is higher than that of the ground state. The curvature of this spectral line is also different from those seen in Fig.~\ref{Fig:ResonantCase}. (2) With increasing coupling strength, the locations of the avoided crossings approach the center, such that the avoided crossings occur at a value of $|\epsilon|$ that is smaller than $\hbar\omega$. This feature can be understood by noting that the cavity field that forms at finite $\epsilon$ acts back on the qubits and makes an additional contribution to $\epsilon$ such that the net value of $\epsilon$ felt by the qubits is larger than the externally applied value. (3) Ignoring the complication of the inward-shifting avoided crossing point mentioned in (2), the spectral signature of the superradiant phase, which is seen at $\epsilon=0$, is generally independent of the avoided crossings at $|\epsilon|\approx\hbar\omega$. Specifically, it is possible that one of these two features can be observable while the other is not, depending on the system parameters. Additional plots that summarize the spectral features expected in the superradiant phase with $\Delta\ll\hbar\omega$ are given in the Appendix. When compared to the case $\Delta=\hbar\omega$, it is clear that in the case $\Delta\ll\hbar\omega$ we have a smaller separation between the two spectral lines in the upper branch in the superradiant phase, meaning that the experimental observation might be more difficult. However, since these figures are all plotted in units of $\omega$, which is quite large, the experimental observation might also be possible in this case.

It should also be noted that in Fig.~\ref{Fig:SmallDeltaCase}, as well as in other figures, the nature of the excitations of any given spectral line generally changes as $\epsilon$ changes. For example, in Fig.~\ref{Fig:SmallDeltaCase}(c) the cavity excitation mode is described by the spectral line with frequency close to $\omega$. This association between the spectral lines and the cavity excitation mode changes depending on the value of $\epsilon$. As a result, if the probe used to obtain the spectrum couples to the cavity but not to the qubits, the observed spectrum will typically contain only parts of each spectral line in Fig.~\ref{Fig:SmallDeltaCase}(c). In particular, one will typically observe three spectral lines that exhibit avoided crossings at regions where one of the lines gradually fades out and the next line starts as one moves along the $\epsilon$ axis. On the other hand, the parts of the spectra lines that have frequencies far from $\omega$ might not be observed in the experiment, because these parts of the spectral lines correspond to exciting the qubit mode.

\section{Effect of fluctuations in system parameters; Spectral line widths}
\label{Sec:ParameterFluctuations}

We now consider how the situation changes when we include fluctuations in the parameters $\Delta$, $\epsilon$ and $g$. In particular, we are interested in how these fluctuations will affect the observed spectral lines. Before we start the discussion of the case of non-identical qubits we note that the case of identical qubit parameters allowed us to perform rigorous derivations for the spectra, while here we will have to rely on some qualitative arguments. We also note that a related study on the effect of parameter fluctuations on superradiance in superconducting circuits was performed in Ref.~\cite{Lambert}.

The ground state in this case has been described in Ref.~\cite{AshhabSemba}. The electromagnetic field of the resonator has an average value that is determined by the many contributions from all the interactions with the individual qubits in the ensemble, and the state of each qubit is determined by its bias parameter including a contribution from the possibly finite value of the resonator's electromagnetic field. If we consider exciting the qubit modes, we have to take into account the fact that the qubits now do not behave collectively as above (at least not having sharp spectral features as in the case of identical qubits). Let us for a moment consider a qubit ensemble that is not coupled to the resonator. We can analyze the excitation modes of the ensemble by considering the large collection of possible excitations of the individual qubits. In other words, each qubit has its own values of $\Delta$ and $\epsilon$, giving a hyperbolic spectrum, and the spectra of the individual qubits are then superimposed to produce the spectrum of the ensemble. As a result, if the parameters of the individual qubits have large variations, the spectral line for the qubit ensemble can be very broad. This broadening can make it more difficult to measure fine features in the spectrum, such as the appearance of additional spectral lines in the superradiant phase, if the separation between the spectral lines is smaller than their widths. The cavity's oscillation mode, on the other hand, should exhibit a different behavior. The picture used in the previous sections (and in Ref.~\cite{AshhabSemba}) leads to the conclusion that the single cavity mode feels the mean-field force applied by the qubit ensemble, and as a result its resonance frequency (although possibly strongly modified by the interaction with the qubit ensemble) should remain well defined and not drastically broadened by the broad qubit frequency distribution. Indeed, the spectra in Ref.~\cite{Kakuyanagi} have a width that is broadened by a factor of 2 at the symmetry point compared to the width far away from the symmetry point, which shows that the large qubit parameter fluctuations do not significantly broaden the oscillator's spectral line.

One possible complication occurs when a significant fraction of the qubit frequencies are resonant with the cavity frequency (which, as mentioned above, can be different from the bare cavity frequency because of renormalization). This situation occurs for example at the avoided crossings in Fig.~\ref{Fig:SmallDeltaCase}(a-c). There are two competing mechanisms at play at such avoided crossings. On one hand, if we consider the case of no parameter fluctuations, the minimum gap of the avoided crossing is approximately given by $g_{\rm eff}=\sqrt{N}g\cos\phi$, where $\tan\phi=\epsilon/\Delta$. On the other hand, assuming that all qubit gaps $\Delta_i$ are much smaller than $\hbar\omega$, the qubit frequencies have a spread given by the width of the distribution of $\epsilon_i$ values, which we call $\delta\epsilon$. If $g_{\rm eff}\gg\delta\epsilon$, one would expect the avoided crossing to be observed with the width $\delta\epsilon$ determining the spectral line widths. If on the other hand $g_{\rm eff}\ll\delta\epsilon$, one would expect that no (clear) avoided crossing will be observed. Considering this situation in detail, one can argue as follows: the region in which the qubit and cavity frequencies cross will be broad because of the broad distribution of $\epsilon_i$. At any given point in this region, the qubits with frequencies that are closer to the cavity frequency than the qubit-cavity coupling strength will experience avoided crossings with the cavity. In some sense, the coupling between the cavity and each one of the resonant qubits gives rise to an avoided crossing whose size is proportional to the coupling strength. When instead of one avoided crossing we have a large number of closely spaced avoided crossings, the many closely spaced spectral lines merge and we effectively obtain a single broadened spectral line. As a result the cavity's spectral line will have a width that is given by the qubit-oscillator coupling strength with some partial ensemble enhancement, but this width can be significantly smaller than the width of the whole qubit ensemble frequency distribution. 

We therefore expect that the conditions needed to observe the features related to the superradiant phase and the conditions needed to observe the features related to coherent resonant coupling between the qubit ensemble and the cavity to be different. The fact that these are two different and somewhat independent phenomena can be seen in Fig.~\ref{Fig:SmallDeltaCase}(c). To observe the signature of superradiance, one can look for the spectral line shown by the magenta dashed line near $\epsilon=0$. In the superradiant phase, one expects to find a range of $\epsilon$ values where two spectral lines instead of one appear near the cavity's resonance frequency, and these lines cross to give an X shape [which is more clearly seen in other figures, e.g.~Fig.~\ref{Fig:SmallDeltaCase}(d)]. The widths of these lines is expected to be roughly the width of the cavity's spectral line, which can be much narrower than the qubit frequency distribution. To observe coherent resonant coupling, one would look for the avoided crossings around $|\epsilon|\approx\hbar\omega$. The condition to observe this avoided crossing is $\sqrt{N}g\overline{\Delta}/\sqrt{\overline{\Delta}^2+\overline{\epsilon}^2}\gg\delta\epsilon$, where $\overline{\Delta}$ and $\overline{\epsilon}$ are the average values of $\Delta$ and $\epsilon$, and $\delta\epsilon$ is the standard deviation in the values of $\epsilon$ at the avoided crossing.

\section{Experimental considerations}
\label{Sec:ExperimentalConsiderations}

As a reference point for the discussion in this section, we consider the experimental setup of Ref.~\cite{Kakuyanagi}. The qubit ensemble has $\sim$4000 qubits with gaps $\sim$1 GHz, while the resonator frequency is $\sim$6 GHz. The coupling strength between a single qubit and the resonator is $\sim$15 MHz. This combination of parameters gives $4g^2N/(\hbar\omega\Delta)\sim 0.6$, which corresponds to the normal phase but is close to the critical point. For example, if we take the value $g\sim$ 25-30 MHz, which gives $4g^2N/(\hbar\omega\Delta)\sim$ 1.6-2.4, the system parameters will be well inside the superradiance regime. The metastable excited state will then exist up to values of $|\epsilon|/\omega\sim$ 0.1-1, which would be easily accessible experimentally. Unless $4g^2N/(\hbar\omega\Delta)\approx 1.7$, the separation between the two spectral lines in the high-frequency modes will be at least a few percent of $\omega$, i.e.~at least tens of MHz.

The lifetime of the metastable state can be estimated by considering that the energy barrier between the two stable states at the symmetry point is $(Ng)^2/(\hbar\omega)$. Transforming the full quantum state from one of these many-body states to the other would require flipping the states of $N$ qubits. Since the flipping of qubit states in the basis of $\sigma_z$ states is induced by the $\hat{\sigma}_x$ operator, each flip of a single-qubit state when moving between the ground and metastable states is associated with a transition matrix element on the order of $\Delta \exp[-2g^2/(\hbar\omega)]$. Combining these estimates we obtain a rough estimate for the metastable state's decay rate given by $\Delta [\hbar\omega\Delta/(Ng)^2]^N \exp[-2Ng^2/(\hbar\omega)]$. For parameters that are well into the superradiance regime (e.g.~$g/g_c>1.5$) and more than a few qubits, we find that the metastable state's decay rate close to $\epsilon=0$ is orders of magnitude smaller than $\Delta$ and decreases exponentially with increasing $N$. Coupling to environmental degrees of freedom can suppress this decay rate further, as occurs in the spin-boson model \cite{Leggett}. When the system is biased away from the symmetry, the lifetime of the metastable excited state decreases gradually and eventually goes down to zero as $\epsilon$ approaches the point where the double-well potential of Fig.~\ref{Fig:DoubleWellPotential} turns into a single well. As a result and as mentioned above, the state will be short lived close to that instability point. One should therefore look for the signatures of the superradiant phase close to the symmetry point.

Assuming that the metastable excited state is sufficiently long lived, it can be accessed by first setting the bias parameter $\epsilon$ to a large value such that no metastable excited state exists and then sweeping $\epsilon$ across the symmetry point. This way, what used to be the ground state adiabatically evolves into the metastable excited state. Then the observed spectrum will be that of the metastable excited state until it decays to the ground state.

We finally note that to obtain the spectra discussed in this paper, no coherent superposition is required between the macroscopically different ground and metastable states, even at the symmetry point. Realizing such a macroscopic superposition involving $\sim$4000 qubits and many-photon coherent states in the resonator would be extremely challenging with currently available superconducting circuits. The recent experiment in Ref.~\cite{YoshiharaDSC} demonstrated evidence of highly entangled energy eigenstates in a single-qubit-single-oscillator circuit. However, even in that case, the energy separation ($\sim$10-100 MHz) between the lowest two energy eigenstates was 1-2 orders of magnitude smaller than the energy scale set by the temperature ($\sim$40 mK). As a result, the thermal equilibrium entanglement was estimated to be on the order of a few percent.

\section{Conclusion}
\label{Sec:Conclusion}

We have analyzed the spectra that one expects to observe in a superconducting realization of the Dicke model, with a qubit ensemble coupled to a single harmonic oscillator. We have identified the appearance of additional spectral lines as a possible signature of the superradiant phase. Depending on the specifics of the experimental setup, such as the measurable frequency range and the lifetime of the metastable excited state, different features in the spectrum can be easier to observe than others. For typical parameters of superconducting circuits, we expect that one will be able to observe some signature of the superradiant phase if the coupling strength exceeds the critical value.

\section*{Acknowledgment}

This work was supported in part by Japan Science and Technology Agency (JST) Core Research for Evolutionary Science and Technology (CREST) Grant Numbers JPMJCR1774 and JPMJCR1775, Japan.

\section*{Appendix: additional plots}

In this Appendix, we show additional plots that supplement those shown in the main text.

\begin{center}
\textbf{Resonant case}
\end{center}

\begin{figure}[h]
\includegraphics[width=8.0cm]{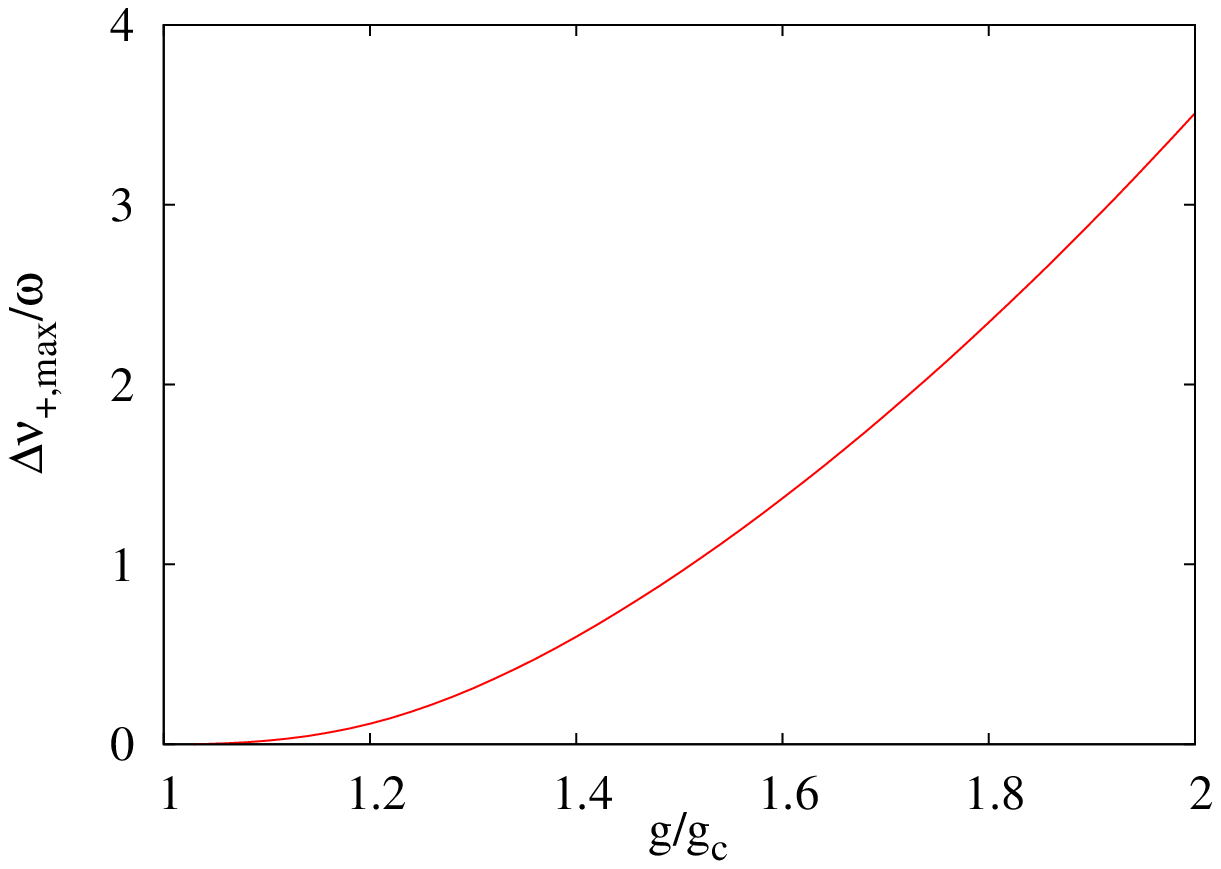}
\caption{The maximum frequency separation between the two $\nu_+$ spectral lines (i.e.~at the maximum value of $|\epsilon|$ before one of them disappears) as a function of $g/g_c$. Here we take the resonant case $\Delta=\hbar\omega$.}
\label{Fig:MaximumLineSeparationResonant}
\end{figure}

\begin{figure}[h]
\includegraphics[width=8.0cm]{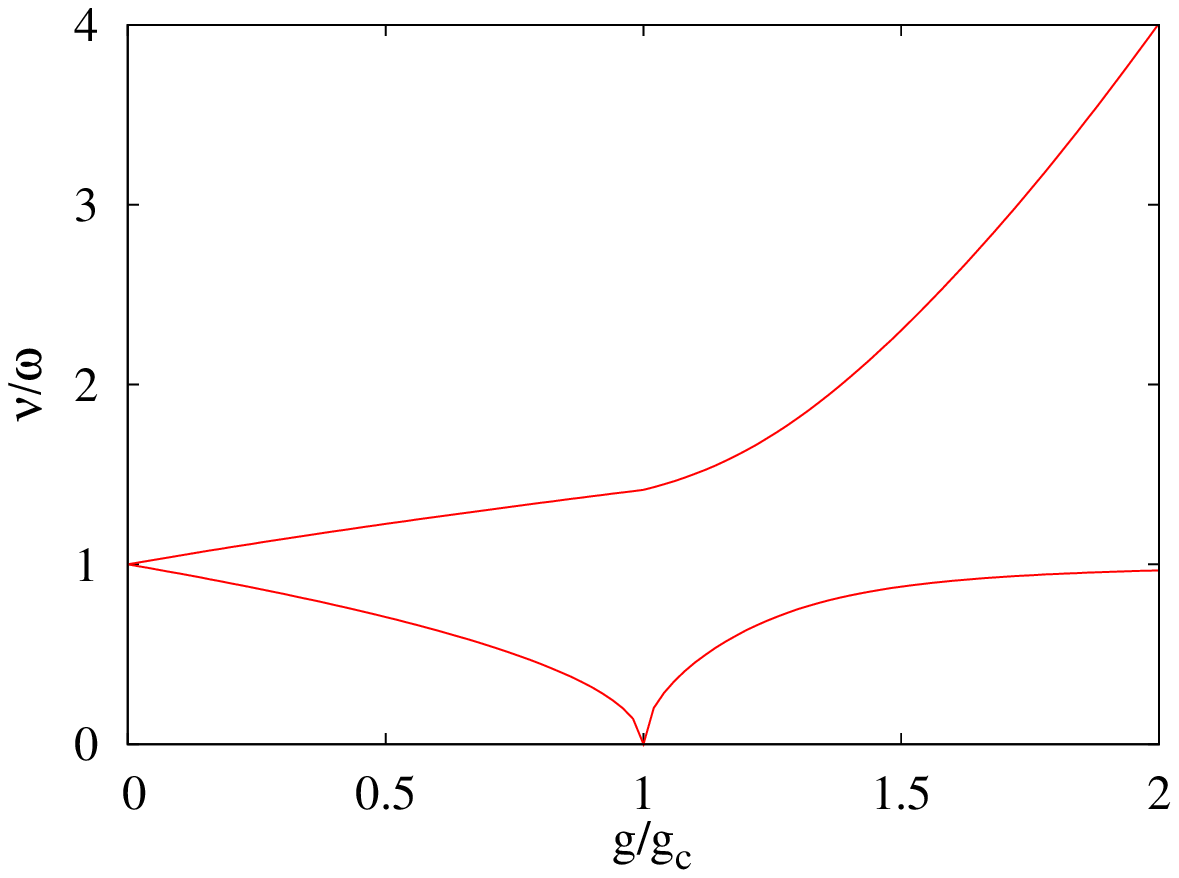}
\caption{The frequencies $\nu_{\pm}$ at the symmetry point as functions of $g/g_c)$ for the case $\Delta=\hbar\omega$, where $g_c=\hbar\omega/(2\sqrt{N})$.}
\label{Fig:FrequenciesAtSymmetryPointResonant}
\end{figure}

The maximum separation between the two spectral lines in the upper branch, which occurs at the maximum value of $|\epsilon|$ before one of the two lines disappears (i.e.~the value of $|\epsilon|$ plotted in Fig.~4 in the main text) is plotted in Fig.~\ref{Fig:MaximumLineSeparationResonant}. To give an idea about the frequency ranges where the above spectral features are expected to be found, Fig.~\ref{Fig:FrequenciesAtSymmetryPointResonant} shows the two frequencies $\nu_{\pm}$ at the symmetry point as functions of coupling strength. It is worth mentioning that the frequencies shown in Fig.~\ref{Fig:FrequenciesAtSymmetryPointResonant} are well known in the literature (see e.g.~Ref.~\cite{EmaryPRE}).

\begin{center}
\textbf{Non-resonant case}
\end{center}

\begin{figure}[h]
\includegraphics[width=6.0cm]{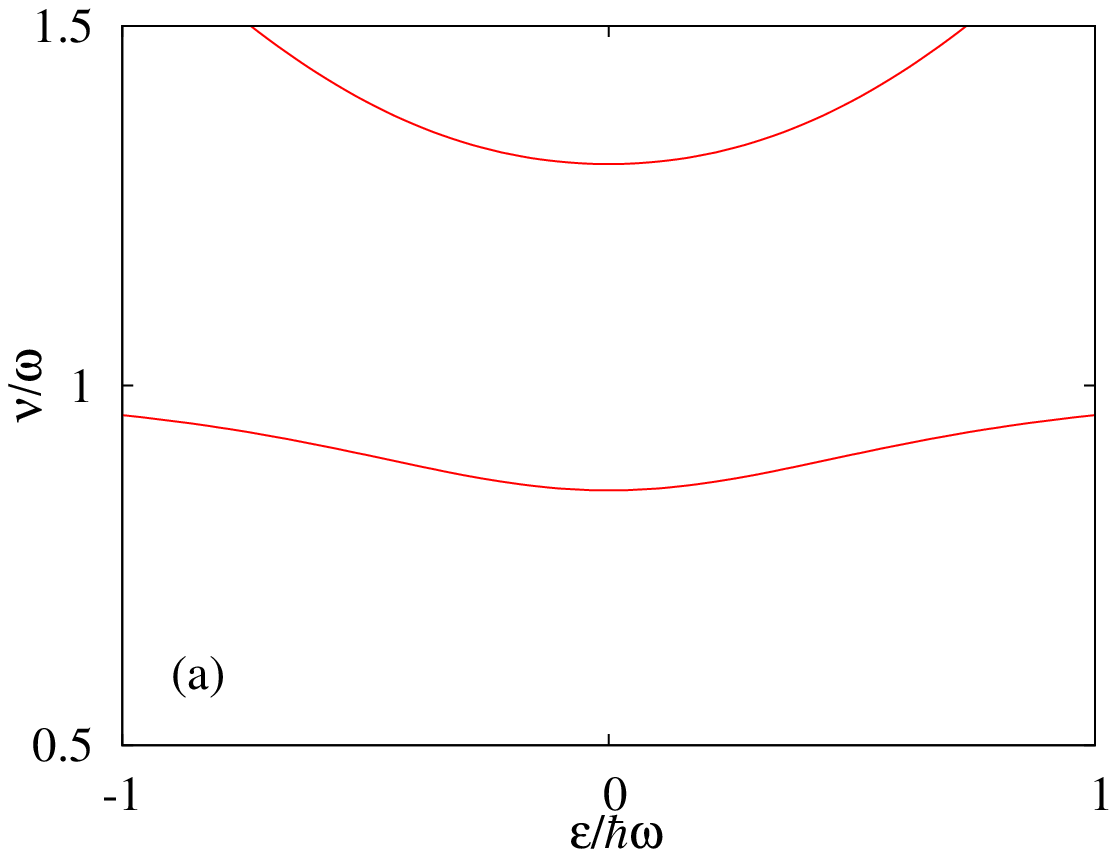}
\includegraphics[width=6.0cm]{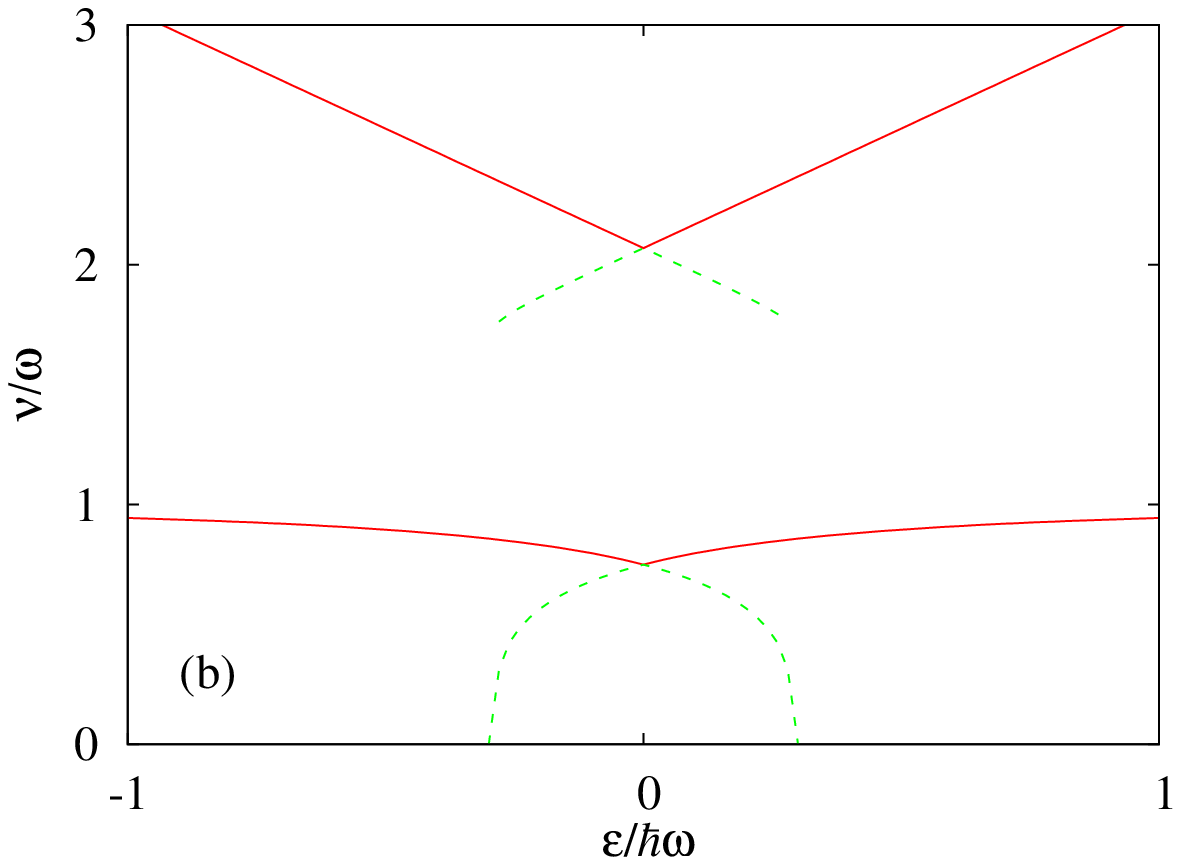}
\includegraphics[width=6.0cm]{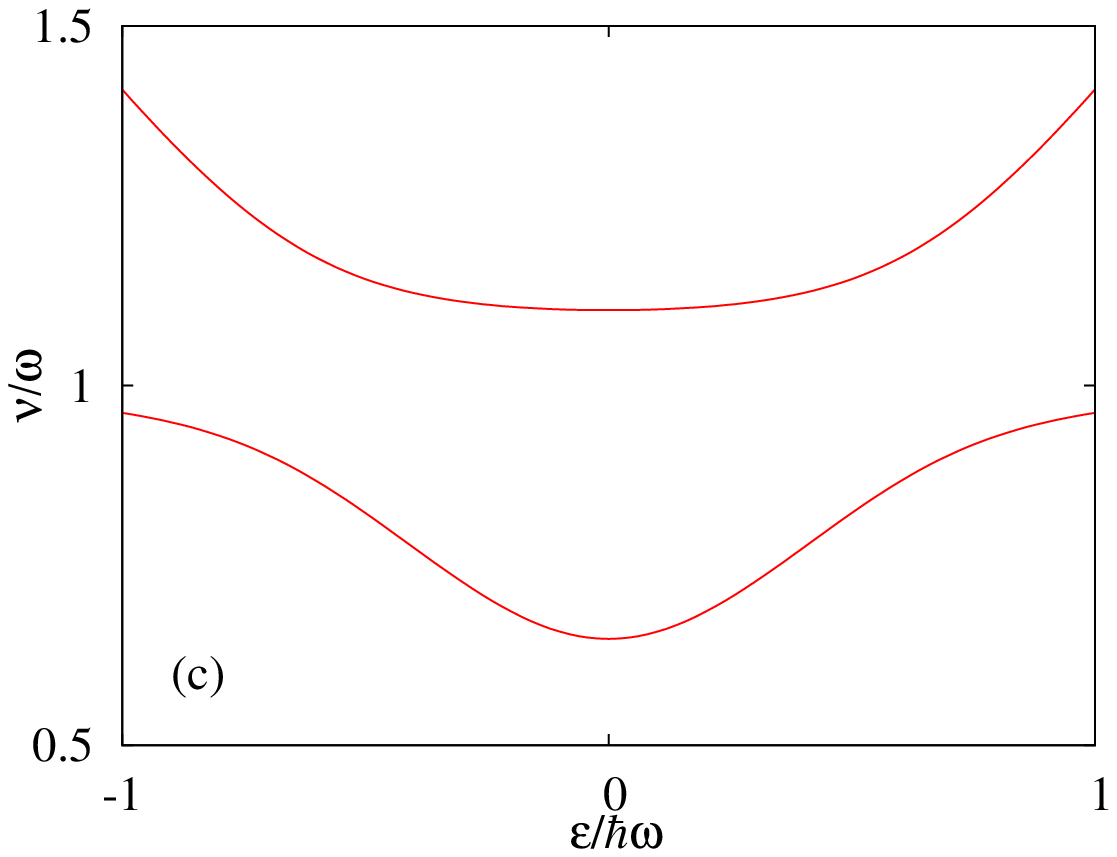}
\includegraphics[width=6.0cm]{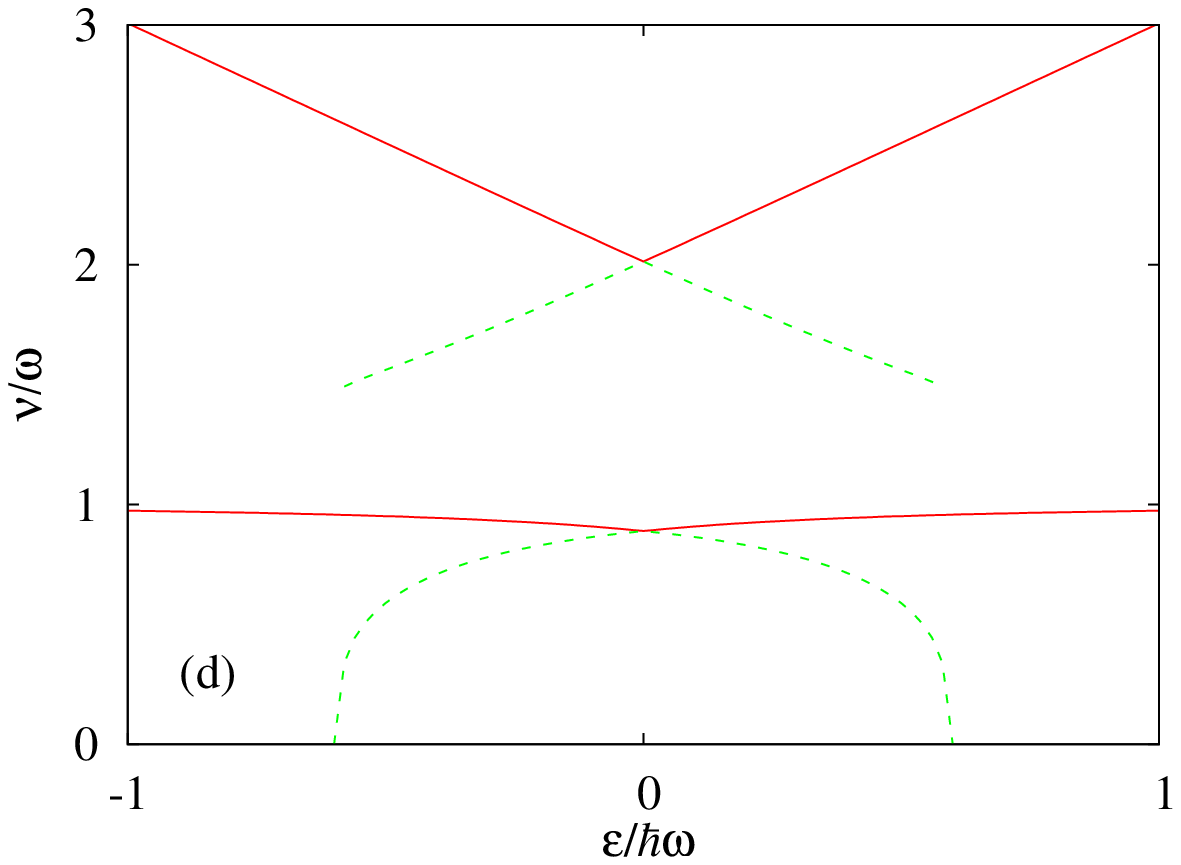}
\caption{Frequencies of the excitation modes $\nu_{\pm}$ as functions of $\epsilon/(\hbar\omega)$. As in Fig.~2 in the main text, the red solid lines are the spectral lines that correspond to excitation from the ground state, while the green dashed lines correspond to excitation from the metastable excited state. Here we take the non-resonant case $\Delta\neq\hbar\omega$. In (a) we set $\Delta/(\hbar\omega)=1.2$, $\sqrt{N}g/(\hbar\omega)=0.2$. In (b) we set $\Delta/(\hbar\omega)=1.2$, $\sqrt{N}g/(\hbar\omega)=0.7$. In (c) we set $\Delta/(\hbar\omega)=0.8$, $\sqrt{N}g=0.2$. In (d) we set $\Delta/(\hbar\omega)=0.8$, $\sqrt{N}g/(\hbar\omega)=0.7$. The spectra are qualitatively similar to those shown in Fig.~2 in the main text.}
\label{Fig:NonResonantCase}
\end{figure}

Figure \ref{Fig:NonResonantCase} shows spectra for two cases with $\Delta\neq\hbar\omega$. The spectra look generally similar to those shown in Fig.~2 in the main text, especially in the superrandiant phase.

\begin{center}
\textbf{Small qubit gap}
\end{center}

\begin{figure}[h]
\includegraphics[width=8.0cm]{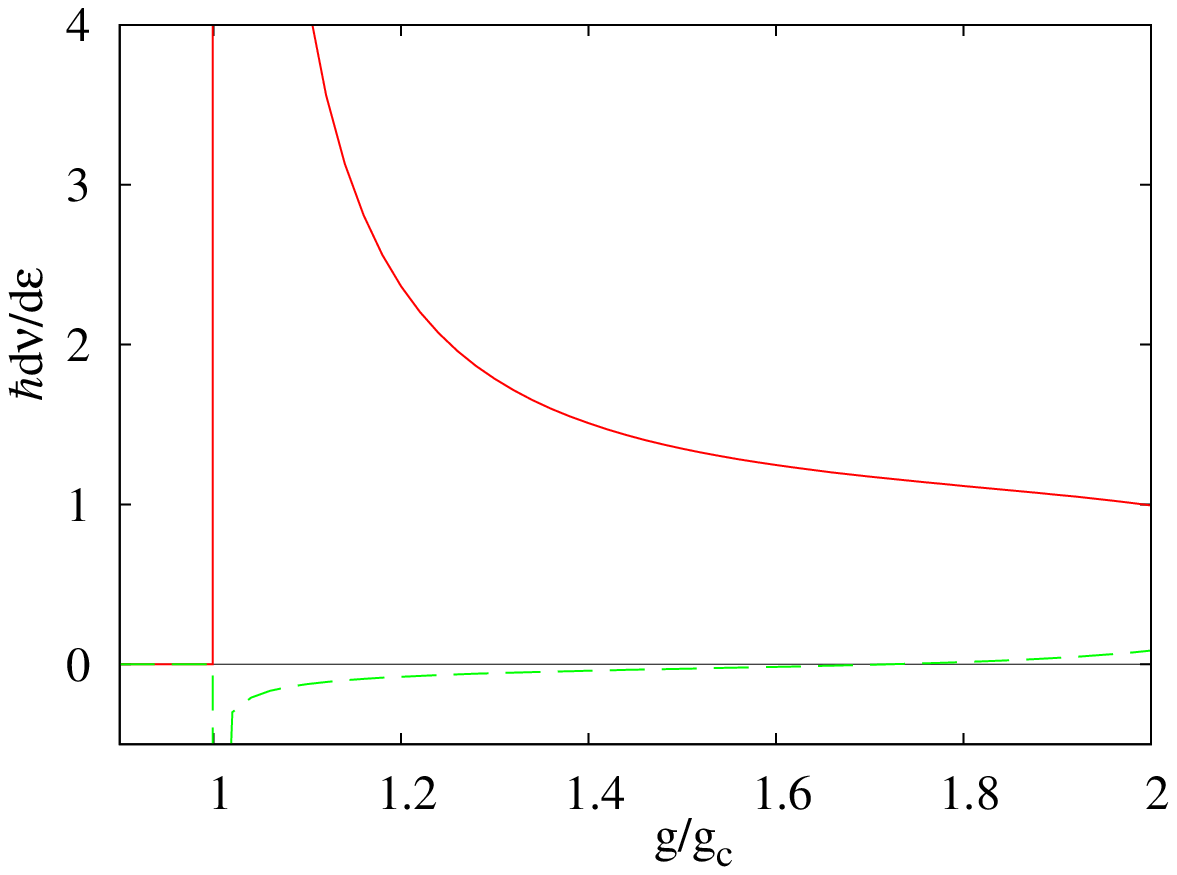}
\caption{Slopes of the frequencies $\nu_-$ (red solid line) and $\nu_+$ (green dashed line) with respect to $\epsilon$ (i.e.~$\hbar d\nu_{\pm}/d\epsilon$) at the symmetry point as a function of $g/g_c$. As in Fig.~5 in the main text, we set $\Delta/(\hbar\omega)=0.2$.}
\label{Fig:SlopeAtSymmetryPointLowFreqQubits}
\end{figure}

\begin{figure}[h]
\includegraphics[width=8.0cm]{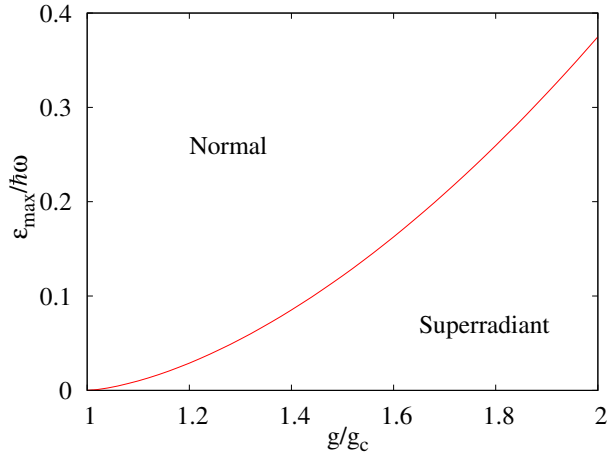}
\caption{The maximum value of $|\epsilon|$ at which there exists a metastable state, i.e.~the maximum value of $|\epsilon|$ at which one expects to see four lines in the spectrum, as a function of $g/g_c$. As above $\Delta/(\hbar\omega)=0.2$. The labels ``Normal" and ``Superradiant" describe the phase that is obtained in each region in the $g$-$\epsilon$ parameter space.}
\label{Fig:BistabilityRangeLowFreqQubits}
\end{figure}

\begin{figure}[h]
\includegraphics[width=8.0cm]{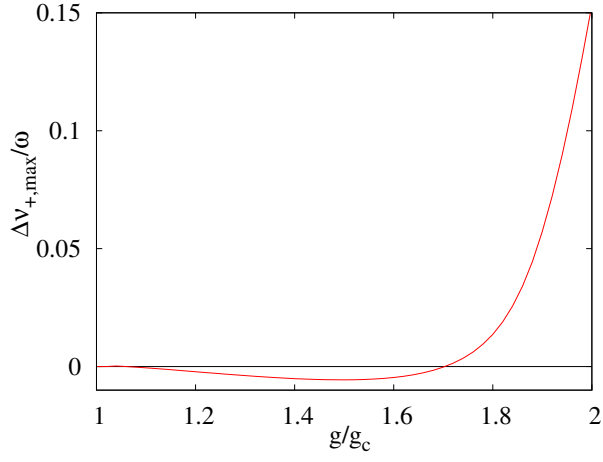}
\caption{The frequency separation between the two $\nu_+$ spectral lines at the maximum value of $|\epsilon|$ before one of them disappears as a function of $g/g_c$. The fact that the difference is negative just above $g_c$ means that the spectral lines have a shape resembling the letter W near the symmetry point, as can be seen in Fig.~5(c) in the main text. As above $\Delta/(\hbar\omega)=0.2$.}
\label{Fig:MaximumLineSeparationLowFreqQubits}
\end{figure}

\begin{figure}[h]
\includegraphics[width=8.0cm]{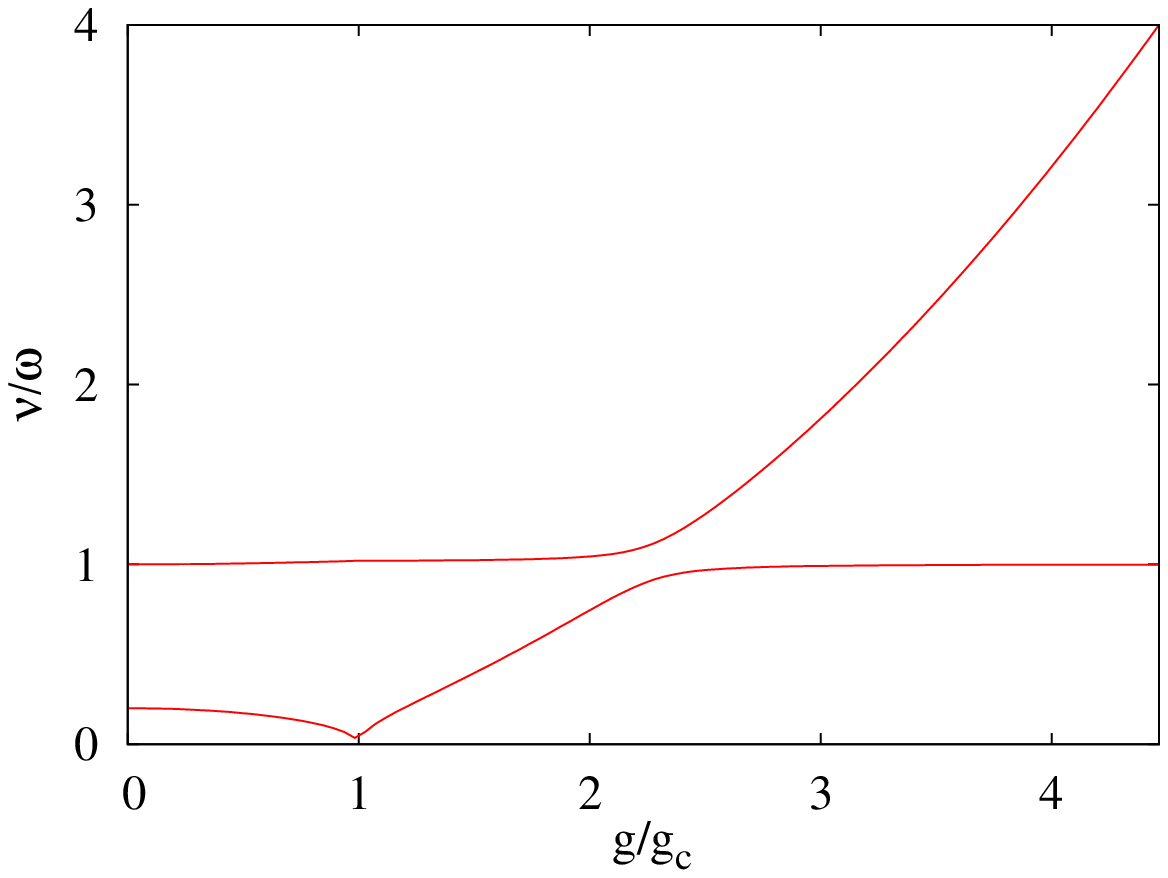}
\caption{The frequencies $\nu_{\pm}$ at the symmetry point as functions of $g/g_c$ for the case $\Delta/(\hbar\omega)=0.2$, where $g_c=0.224\hbar\omega/\sqrt{N}$.}
\label{Fig:FrequenciesAtSymmetryPointLowFreqQubits}
\end{figure}

Figures \ref{Fig:SlopeAtSymmetryPointLowFreqQubits}-\ref{Fig:MaximumLineSeparationLowFreqQubits} summarize the spectral features expected in the superradiant phase with $\Delta\ll\hbar\omega$. In Fig.~\ref{Fig:SlopeAtSymmetryPointLowFreqQubits} we plot the slopes of both the spectral lines at the symmetry point. One obvious difference that we can see between Fig.~3 in the main text and Fig.~\ref{Fig:SlopeAtSymmetryPointLowFreqQubits} is that the slope of the $\nu_+$ line in Fig.~\ref{Fig:SlopeAtSymmetryPointLowFreqQubits} is negative just above the critical coupling strength. The $\nu_-$ spectral line exhibits qualitatively similar behavior in the two cases. In Fig.~\ref{Fig:BistabilityRangeLowFreqQubits} we plot the range of $\epsilon$ values for which the four spectral lines exist. The behavior is qualitatively similar to that shown in Fig.~4 in the main text. The maximum separation between the two spectral lines in the upper branch is plotted in Fig.~\ref{Fig:MaximumLineSeparationLowFreqQubits}. In Fig.~\ref{Fig:FrequenciesAtSymmetryPointLowFreqQubits} we plot the two frequencies $\nu_{\pm}$ at the symmetry point as functions of coupling strength.

\end{document}